\documentclass[fleqn,usenatbib]{mnras}

\usepackage{newtxtext,newtxmath}

\usepackage{caption}
\usepackage{subcaption}

\usepackage[T1]{fontenc}

\DeclareRobustCommand{\VAN}[3]{#2}
\let\VANthebibliography\thebibliography
\def\thebibliography{\DeclareRobustCommand{\VAN}[3]{##3}\VANthebibliography}

\usepackage{graphicx}	
\usepackage{amsmath}	

\usepackage{placeins}
\usepackage{orcidlink}

\title[Supermassive stars due to internal Lyman-Werner feedback]{Can supermassive stars form in protogalaxies due to internal Lyman-Werner feedback?}

\author[J. Sullivan et al.]{
James Sullivan$^{1}$\thanks{E-mail: jms2561@columbia.edu}\orcidlink{0009-0008-0904-5535},
Zolt\'an Haiman$^{1,2,3}$\orcidlink{0000-0003-3633-5403},
Mihir Kulkarni$^{4,5}$\orcidlink{0000-0002-9789-6653},
Eli Visbal$^{4}$\orcidlink{0000-0002-8365-0337}
\\
$^{1}$Department of Astronomy, Columbia University, 550 West 120th Street, New York, NY, 10027, USA\\
$^{2}$Department of Physics, Columbia University, 550 West 120th Street, New York, NY, 10027, USA\\
$^{3}$Institute of Science and Technology Austria (ISTA), Am Campus 1, Klosterneuburg 3400, Austria\\
$^{4}$University of Toledo, Department of Physics and Astronomy and Ritter Astrophysical Research Center, 2801 W. Bancroft Street,
Toledo, Ohio 43606\\
$^{5}$Institut für Astrophysik, Georg-August Universität Göttingen, Friedrich-Hund-Platz 1, D-37077 Göttingen, Germany
}

\date{Accepted XXX. Received YYY; in original form ZZZ}

\pubyear{2025}

\begin{document}
\label{firstpage}
\pagerange{\pageref{firstpage}--\pageref{lastpage}}
\maketitle

\begin{abstract}
Population III stars are possible precursors to early massive and supermassive black holes (BHs). The presence of soft UV Lyman Werner (LW) background radiation can suppress Population III star formation in minihalos and allow them to form in pristine atomic cooling halos. In the absence of molecular hydrogen ($\rm H_2$) cooling, atomic-cooling halos enable rapid collapse with suppressed fragmentation. High background LW fluxes from preceding star-formation have been proposed to dissociate $\rm H_2$. This flux can be supplemented by LW radiation from one or more Population III star(s) in the same halo, reducing the necessary background level. Here we consider atomic-cooling halos in which multiple protostellar cores form close to one another nearly simultaneously. We assess whether the first star's LW radiation can dissociate nearby $\rm H_2$, enabling the prompt formation of a second, supermassive star (SMS) from warm, atomically-cooled gas. We use a set of hydrodynamical simulations with the code \texttt{ENZO}, with identical LW backgrounds centered on a halo with two adjacent collapsing gas clumps. When an additional large local LW flux is introduced, we observe immediate reductions in both the accretion rates and the stellar masses that form within these clumps. While the LW flux reduces the $\text{H}_2$ fraction and increases the gas temperature, the halo core's potential well is too shallow to promptly heat the gas to $\gtrsim$ 1000 K and increase the accretion rate onto the second protostar. We conclude that internal LW feedback inside atomic-cooling halos is unlikely to facilitate the formation of SMSs or massive BH seeds.
\end{abstract}




\section{Introduction}
\label{sec:Intro}
Supermassive black holes (SMBHs) are found at the center of nearly all nearby galaxies \citep{Kormendy13} and play a major role in their evolution. However, the formation and evolution of these SMBHs is still shrouded in uncertainty. There are several proposed routes to explain the presence of SMBHs as massive as $10^{8-9} \, \rm M_{\odot}$ as early as $ z \sim 6$ \citep{Fan06, Venemans15,Wu_2015, Banados18, Inayoshi20, Wang21, Fan23, Bosman24}. The most common pathways can be divided into the following categories: a ``normal'' massive Population III (hereafter Pop III) star ($\sim$$10^{1-2}\, \rm M_\odot$) forms and its remnant BH accretes large amounts of mass, possibly at hyper-Eddington rates, to form a SMBH; an intermediate-mass BH (IMBH; $\sim$$10^{3-4}\, \rm M_\odot$) forms and grows to a SMBH through accretion and/or mergers; or finally, a massive primordial star ($\sim$$10^{5-6}\, \rm M_\odot$) forms, collapses promptly into a massive BH, and then grows by steady accretion \citep{Inayoshi20,Volonteri21}. 

Several cosmological simulations have found inefficient growth of stellar-mass BH seeds, making the first route disfavoured
\citep{Alvarez09, Milo09, Smith18,Spinoso22}. However, rare periods of high, tumultuous accretion could provide a pathway to large masses for a small subset of these stellar-mass seeds. If the direction of these accretion flows is uncorrelated, the BH will have lower spin rates and radiative efficiency, consequently increasing growth ~\citep{Zubovas21}.

Forming SMBHs from intermediate-mass seeds requires runaway collisions and is therefore limited to dense stellar clusters. However, studies demonstrate difficulties growing black holes larger than $10^{3-4} \, \rm M_\odot$ \citep{Zwart04, Omukai08, Devecch09, Katz15, Gonzalez21, Rizzuto21}. Supermassive stars have consequently been proposed as a route to form SMBHs, avoiding the need for high collision rates or super-Eddington accretion \citep{Begelman78, Rees78, Begelman06, Begelman08, Latif16, Woods19}. In this paper, we focus on this latter scenario.


Simulations predict that Pop III stars form in $10^5-10^6 \, \rm M_\odot$ minihalos \citep{Haiman96, Tegmark97, Machacek01, Abel02, Bromm02, Yoshida03, Hirano14, Greif15, Kulkarni21, Klessen23}. These first stars form in metal-free regions and are composed of hydrogen, helium, and trace amounts of lithium \citep{Yoshida12, Bromm13, Klessen19}. These pristine halos have virial temperatures less than the atomic-cooling threshold, $\sim10^4 \, \rm K$, so they cool via molecular hydrogen ($\text{H}_2$) ro-vobrational transition lines \citep{Bromm03, Bromm11}. However, supernovae quickly enrich the gas with metals. This is in part due to the short stellar lifespans of massive Pop III stars, which are on the order of a few Myr. The higher efficiency of metal cooling leads to transition to lower-mass Population II (Pop II) stars. 

Pop III stars produce large numbers of Lyman Werner (LW) photons, ranging from $11.2$ to $13.6 \rm \,eV$ and above, throughout their lifetime. These contribute to an early background LW radiation as the star formation density increases. This radiation can drastically affect future star formation. In pristine halos, the LW radiation can dissociate $\text{H}_2$ and prevent cooling by its rovibrational transitions. Thus in regions with strong LW radiation, Pop III star formation will be suppressed in minihalos with virial temperatures $T_\text{vir} \lesssim 10^4 \, \rm K$, which rely on  $\text{H}_2$ to cool \citep{Haiman96, Tegmark97, Haiman00, Machacek01, OShea07, Wise07}. 

Atomic-cooling halos (ACH) are thus a promising potential location for massive Pop III star formation. These halos have total (dark matter + gas) masses of $10^{7-8} \, \rm M_\odot$, corresponding to virial temperatures of $T_{\text{vir}} \approx 10^4 \, \rm K$ at redshifts $10-20$ \citep{Omukai01, Prieto13, Becerra14, Regan20}. In the absence of LW radiation, collapsing gas experiences rapid $\rm H_2$ cooling and fragments. This favors the formation of multiple ``normal'' Pop III stars at the Jeans mass corresponding to the temperature at which cooling becomes inefficient (see review by \citealt{Inayoshi20}). However, with intense LW radiation, $\rm H_2$ is dissociated and the temperature remains roughly isothermal at $T \simeq 10^4 \, \rm K$ for much of the collapse. With the corresponding large gas accretion rate, a supermassive star can form with a mass as high as $10^{5-6} \, \rm M_\odot$ \citep{Bromm02, Wise08, Regan09b, Regan09a}. Recent studies \citep{Hosokawa12, Hosokawa13b, Woods17, Haemmerle18,Nandal23} have set this critical accretion rate between $0.01-0.04 \, \rm M_\odot \,yr^{-1}$. 2D simulations by \cite{Sakurai16} showed that brief periods of slower accretion are allowed and will not prevent SMS formation, as long as the duration of such an episode does not exceed $10^3 [M_*/500M_\odot]^\frac{1}{2}\,\rm yr$. In the absence of this rapid accretion, the star will settle onto the main sequence at the lower mass of a ``normal'' massive Pop III star of order $\sim100~{\rm M_\odot}$ and limits its own growth via its UV radiation. 

Again, this route requires a strong LW flux to dissociate any $\rm H_2$. The critical flux, $J_{\rm crit}$ is set approximately by balancing the dissociation rate of $\rm H_2$ with the formation rate (\citealt{Omukai01,Shang10}; see also the review by \citealt{Inayoshi20}). The higher gas densities produced in ACHs increase the $\rm H_2$ formation rate. This consequently raises the critical flux required to destroy this $\rm H_2$ by several orders of magnitude when compared to minihalos. $J_{\rm crit}$ has been estimated to be in the range of $10^{3-5} \, \rm J_{21}$ for ACHs, where $\rm J_{21}$ $ = 10^{-21} \rm \,erg \, s^{-1} \, cm^{-2}\,Hz^{-1}\,sr^{-1}$ \citep{Omukai01,Shang10,Sugimura14, Wolcott17, Wolcott19}.

This large background radiation could be found in rare, overdense regions with bright, nearby galaxies \citep{Dijkstra08}. However, it is larger by about two orders of magnitude than the expected background at the time of reionization \citep{Haiman97, Wise07, OShea08}. In order to produce this level of LW radiation, we propose supplementing the cosmological background LW radiation, or the high LW flux from  neighboring halos, with another source. Previous studies have looked at LW radiation from neighboring \citep{Regan17, Chon17} or merging \citep{Visbal14} halos. {\em Here, we focus on sequential star formation within the cores of individual halos}. We propose that the first star to form within an ACH can produce an additional source of ``internal'' LW radiation. If the first protostellar cloud core fails to form a SMS, it can still produce a sufficiently intense LW radiation to irradiate other protostellar cores collapsing nearby, reducing or eliminating $\rm H_2$-cooling in their vicinity. Here we examine whether this internal LW feedback could then raise the gas temperatures and help produce the aforementioned environment of warm atomic gas, needed to form a SMS. 

Halos with this sort of internal LW feedback across multiple star-forming cores in the same halo have been included in simulations with different focuses. \cite{Dunn18} included the effect of internal LW radiation from star formation on massive direct collapse BH seeds in a cosmological simulation with \texttt{GASOLINE}. They found that the dominant LW sources producing massive black holes often resided a few 100 pc away in the same halo. However, their spatial resolution (few$\times10^{4-5}{\,\rm M_\odot}$ or few 100 pc) did not allow resolving the internal structure or the collapse of protostellar cores in halos. We here follow up on the high-resolution simulations of \cite{Kulkarni19} using the cosmological hydrodynamics code \texttt{ENZO}. They subject three halos to varying amounts of ionizing flux, studying the impact on the halo's collapse and the stellar history. We study the evolution of one of these halos, an ACH with a virial mass of $\sim$$3.2\times 10^8 \, \rm M_\odot$ that produces nearby, sequential star formation. This halo first forms stars well above the atomic-cooling threshold, due to the presence of a strong assumed ionizing radiation background, as described in \S~\ref{sec:SimulationSetup} below. The halo has two primary clumps (see Fig.~\ref{ZoomOutDensTemp} below), which we correspondingly label ``clump A'' and ``clump B.'' We perform a new suite of simulations with \texttt{ENZO}, to track the evolution of the highest mass (at the end of the simulation) star within each clump. We investigate how LW radiation from the first star that forms affects the growth of the stars elsewhere in the halo and whether or not a SMS star can form.

To summarize, in this paper, we look at the effects of adding LW radiation from an internal source on the formation of protostellar cores and their subsequent growth. In \S~\ref{sec:SimulationSetup}, we discuss the setup of our simulations and how we model and track star formation and evolution. In \S~\ref{sec:Analytical}, we present analytical estimates for the propagation of LW radiation-driven ${\rm H_2}$ dissociation vs. the usual atomic HI ionization fronts. In \S~\ref{sec:Results}, we describe the impact of the additional internal LW flux on the formation and growth of protostars in the halo. In \S~\ref{sec:Discussion}, we discuss the implications of our findings for SMS and so-called ``direct-collapse BH (DCBH)'' formation. In \S~\ref{sec:Conclusions} we summarize our results and our main conclusions.

\section{Simulation Setup}
\label{sec:SimulationSetup}

\begin{table}
    \centering

    \begin{tabular}{|c|c|c|}
        \hline
         Run &\textit{``Internal''} $J_\text{LW}$ & $t_\text{delay}$ (yr) \\
        \hline\hline
         Background-only & 0 & n/a \\
         Short-delay & $10^4$ & 20,000 \\
         Long-delay & $10^4$ & $250,000$ \\
         No-cooling & $10^{10}$ & 0 \\

        \hline
    \end{tabular}
    \caption{Parameters of the 4 test runs described in Section~\ref{sec:SimulationSetup}. We list the run names, the intensity of the additional ``internal'' LW flux, and the delay with which we add them after the simulation (re)start.}
    \label{tab:Test Runs}
\end{table}

\begin{figure}
    \centering
    \includegraphics[width=.48\textwidth]{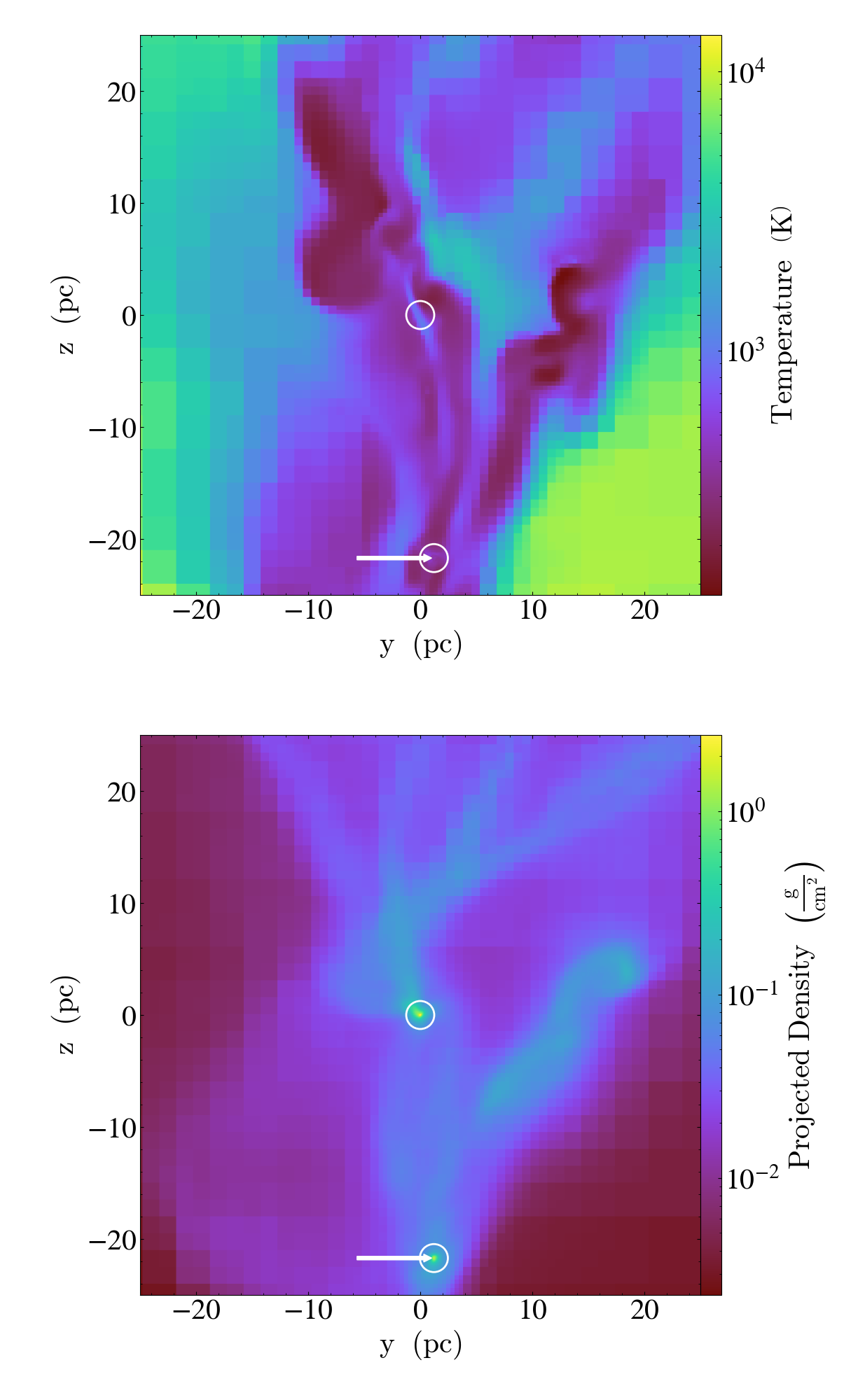}
    \caption{Left: Gas density centered on clump A's maximum density gas cell and projected along the x-axis. Both clump A and clump B are circled. Clump B is marked by the white arrow. Right: Slice of gas temperature along the x-axis and centered on clump A's maximum density gas cell. Both clumps are again circled and an arrow points to clump B. Both the projection and slice are at snapshot \#10, $z = 6.5648$.}
    \label{ZoomOutDensTemp}
    
\end{figure}

We outline our computational and numerical setup below. We first briefly describe \texttt{ENZO}, the hydrodynamical code we use. We then detail our test halo's initial conditions and the numerical setup of our tests. Finally, we specify how we implement LW radiation and measure its effects on sequential star formation within our halo. 
 
We run our simulations using the publicly available adaptive mesh refinement (AMR) code \texttt{ENZO} \citep{Bryan97, OShea04, Bryan14}. \texttt{ENZO} tracks dark matter (DM) dynamics using an N-body particle mesh solver \citep{Shang10} and an Eulerian AMR method produced by \cite{Berger89} to solve the ideal gas hydrodynamic equations. We specifically use the spatially third-order accurate Piecewise Parabolic hydro-solver method, which maintains energy conservation. \texttt{ENZO} self-refines when certain criteria are met, dividing cells into eight smaller cells. This allows it to resolve a wide range of dynamic regions more efficiently. Further, it tracks the  evolution of nine chemical species (H, H+, He, He+, He++, H-, $\rm H_2$+, $\rm H_2$ and e-) and includes radiative cooling \citep{Abel97, Abel00}. We do not include HD or deuterium molecules since these should not affect the regions we are interested in \citep{McGreer08, Kulkarni19}. 

We specifically select Halo C produced in \cite{Kulkarni19} to investigate further. This was one of three halos produced by individual cosmological zoom-in simulations. Each uses a $\Lambda$CDM cosmology agreeing with \cite{Planck14}: $h = 0.67,\, \Omega_b = 0.049,\,  \Omega_m = 0.32, \, \Omega_{\Lambda} = 0.68, \,\sigma_8 = 0.83,$ and $\, n_s = 0.96.$ The simulations were ran in a comoving 2$h^{-1}$Mpc cosmological box centered around each halo. Halo C was selected from a $256^3$ grid DM-only simulation. $128^3$ grid size hydro simulations were then run on each halo. The MUSIC initial conditions generator was set with random seeds that produce matching initial conditions between the $256^3$ and $128^3$ grids. The $128^3$ zoomed hydro simulation had an $836 \, \rm M_\odot$ DM particle mass. The precise cosmological conditions are specified in \cite{Kulkarni19}. We implement the same hydrogen ionization self-shielding prescription, developed by \cite{Rahmati13}, as \cite{Kulkarni19} and \cite{Visbal17}.

Each halo was subjected to a background ionizing flux of $0.1 \, F_0$, where $F_0 = 6.7 \times 10^6 \, \rm photons \, s^{-1} \, cm^{-2}$. This corresponds to the ionizing flux produced by a $6.6 \times 10^{11} \, \rm M_\odot$ dark matter halo at $z \sim7$. This size halo, with the star formation efficiency and escape fraction both 0.1, would produce $2 \times 10^{53} \, \rm{photons} \, s^{-1}$  over a redshift range $\Delta z \simeq 10$. This corresponds to the above $0.1 \, F_0$ for a galaxy 50 kpc away \citep{Kulkarni19}.

The simulation did not use ray-tracing to model the ionizing flux. Instead, it uses an isotropic and uniform radiation field. This is also applied to the ``background'' and ``internal'' LW fluxes we apply. The background LW radiation is set to $100 \, \rm J_{21}$. This is representative of the background LW field in an overdense region of the universe \citep{Ahn09}. This background increases in the presence of outside ionizing radiation as $J_{\text{LW}} = (100 \, + \, 75 \, \times \, F/F_0)\, \rm J_{21}$, with $F$ equaling the ionizing flux  \citep{Kulkarni19}. Our simulation begins at the point Halo C begins runaway collapse, which occurs at $z = 6.5648$. This collapse time is marked by the the simulation reaching refinement level 18. We are therefore modeling a halo whose collapse is delayed due a background ionizing radiation. We note that this atomic cooling halo forms significantly later and is more massive than commonly thought of ACHs. It forms too late for a SMS formed within the halo to grow into a SMBH. However, these facts should not change our general conclusions. We discuss this further in \S~\ref{sec:Conclusions}. 

We use \texttt{ENZO}'s ``sink'' particle mechanism to model star formation and evolution. Sink particles have been previously used to simulate the first generation of stars in grid-based and smoothed-particle hydrodynamics (SPH) codes \citep{Krumholz04, Stacy10, Grief12, Stacy12}. We add a sink particle to a cell when it reaches the highest refinement level and wants to evolve further. The accretion rate is set so that the maximum density of a cell in a spherical region with radius < 5 cell widths cannot exceed the maximum level of refinement. This avoids artificial fragmentation due to forming sinks very close together. We set the maximum refinement level to 18, matching the runs of \cite{Kulkarni19}. Additionally, sink particles merge if the distance between them decreases below 10 times the width of the smallest cell. This sink particle method, used in \cite{Kulkarni19}, produced similar results to more sophisticated sink particle algorithms, i.e. \cite{Regan18}. We are therefore confident in adopting it to this work. We track the sink particle ``properties'' using our data snapshots, which have $\sim$$10^4 \, \rm yr$ temporal resolution.

We include four separate runs in our work (Table~\ref{tab:Test Runs}). Again, each begins at runaway collapse in Halo C from \cite{Kulkarni19}. The first run (``background-only'') proceeds from this point for roughly 1~Myr. We do not include any additional LW flux, setting this up as the benchmark to compare sink particle formation and evolution with and without additional LW radiation. The second run (``short-delay'') implements LW radiation immediately at the simulation (re)start. It ramps up linearly from no additional LW radiation at the simulation restart to $10^4 \, \rm J_{21}$ in 20,000 years. This represents a star ``turning on'' very quickly after the runaway collapse begins within the halo. It then irradiates the halo with photons in the LW band. The third run (``long-delay'') delays adding the additional LW flux by 250,000 years. The flux then increases from the background level to $10^4 \, J_{21}$ at this time. This represents a case in which a protostar forms somewhat later and then eventually settles onto the main sequence before producing LW radiation. The  $10^4 \, \rm  J_{21}$ flux is chosen as the flux since it corresponds to the LW flux produced by a $\sim$$150 \, \rm M_\odot$ Pop III star over its lifetime (several Myr) at a distance of 20 pc away, approximately the distance between our two clumps of gas. 

We find that none of these three initial runs fully dissociate the $\text{H}_2$ in the high density regions surrounding the sink particles. We therefore introduce a much higher (artificial) LW flux of $10^{10} \, \rm J_{21}$ in our fourth run to fully dissociate $\rm H_2$ in the highest density regions and investigate the effect on star formation. This run (``no-cooling'') effectively removes molecular cooling in the halo, and allows us to address whether the gas in the halo is able to dynamically heat to the atomic cooling threshold and activate $\rm HI$ cooling to form a supermassive star.



\FloatBarrier
\section{Estimating the Protostar's Sphere of Influence}
\label{sec:Analytical}
We first analytically treat the problems of $\rm H_2$ dissociation and $\rm H$ ionization within the halo. We estimate the time it would take for a dissociation and ionization front to travel from the center of clump A, host of the first protostar, to the center of clump B, where the second protostar is forming $\sim$$20 \rm \, kpc$ away. We follow the standard $\rm H$-ionization front (e.g. equation~5 of \citealt{Kulkarni19}):
\begin{equation}
   4\pi R^2\,n(R)dR = \left( \dot{N}_{\text{ion}} - 4\pi\alpha\int_{}^{}n^2(r)r^2dr\right)dt,
    \label{eq:Mihir}   
\end{equation}
where R is the radius of spherical shells, $n(R)$ is the hydrogen number density, $\dot{N}_{\text{ion}}$ is the ionizing photon rate, and $\alpha$ is the case-B recombination rate coefficient at $10^4$ K. $r$ is defined as the distance away from the clump center or the sink particle, once it has formed. The number density of hydrogen is measured in the simulations in two regions, 0-0.1 pc and 0.1-20 pc. This covers the distance between clumps A and B in Halo C (Fig.~\ref{ZoomOutDensTemp}).
To calculate the dissociation of molecular hydrogen, we employ the analogous equation:
\begin{equation}
    4\pi R^2\,n_{\text{H}_2}(R)dR = \left( 0.1 \, \dot{N}_{\text{LW}} - 4\pi \,k_9\int_{}^{}{n_{\text{H}}}{n_\text{e}}(r)r^2dr\right)dt,
    \label{eq:dissociation}
\end{equation}
where $k_9$ represents the rate for $\rm H^-$ formation through the combination of $\rm H$ and $\rm e^-$,  $k_9 = 6.775 \times 10^{-15}\; \rm T_{eV}^{0.8779}$, which is the bottleneck reaction in gas-phase ${\rm H_2}$-formation \citep{Shang10}. $n_\text{H}$, $n_{\text{H}_2}$, and $n_\text{e}$ represent the atomic hydrogen, molecular hydrogen, and electron number densities respectively. We multiply the number of LW photon's produced per second, $ \dot{N}_\text{LW}$, by 0.1 because roughly $10$ percent of collisions between LW photons and $\rm H_2$ dissociates the $\rm H_2$.

 We use the hydrogen number density from \cite{Kulkarni19}: $n_0(r/r_\text{pc})^{-2}$, with $r_c =  0.1 \, \rm pc$, for r > 0.1 pc and $n_0$ for r < 0.1 pc, where $n_0 = 1.04 \times 10^{-6}\, \rm cm^{-3}$. The electron number density is produced by creating a spherically averaged profile centered on Clump A. The electron density is then fit logarithmically in the region 0 - 1~pc, and a 3rd-order polynomial in the region 1 - 20~pc. This yields the two expressions:

\begin{equation}
    \rho_e(r) = 
    \begin{cases}
    \small
        \begin{aligned}
            &5.19 \times 10^{-25} \, 
            \exp(-6.42 r_{\mathrm{pc}})\\
            &+ 7.57 \times 10^{-26} \, \mathrm{g \, cm^{-3}} 
        \end{aligned}
        & \text{if } r_{\rm pc} \leq 0.1
        \\[18pt]
        \small
        \begin{aligned}
            &4.542 \times 10^{-30} \, r_{\mathrm{pc}}^4 - 2.131 \times 10^{-28} \, r_{\mathrm{pc}}^3 \\
            &+ 3.503 \times 10^{-27} \, r_{\mathrm{pc}}^2 - 2.397 \times 10^{-26} \, r_{\mathrm{pc}}\\
            & + 1.051 \times 10^{-25} \, \mathrm{g \, cm^{-3}},
        \end{aligned}
        & \text{if } 0.1 \, \leq r_{\rm pc} < 20
    \end{cases}
\end{equation}
where $r_{\rm pc}$ is the distance in parsecs. We model the temperature in both regions with 3rd-order polynomials, producing the two analytical approximations:

\begin{equation}
    T(r) = 
        \begin{cases}
    \small
        \begin{aligned}
            &448.18 r_{\rm pc}^3 + 47.46\,r_{\rm pc}^2 \\ &- 587.33\,r_{\rm pc} + 597.93 \, \rm K
        \end{aligned}
        & \text{if } r_{\rm pc} < 0.1
        \\[18pt]
        \small
        \begin{aligned}
            &6.3571\times10^{-2}\,r_{\rm pc}^3 - 2.52522\,r_{\rm pc}^2 \\ &+ 49.0274 \,r_{\rm pc} + 298.068 \, \rm K
        \end{aligned}
        & \text{if } 0.1 \, < r_{\rm pc} < 20.
    \end{cases}
\end{equation}

These values are used for $k_9$, which is a function of temperature \citep{Shang10}. The flux of LW photons is calculated assuming $6.5 \times 10^4$ LW photons per baryon for a Pop III star \citep{Feathers23}. For a $150 \, \rm M_\odot$ star with a lifetime of 2~Myr, this becomes $1.84 \times 10^{50}$ LW photons per second. This corresponds to $10^4 \rm \, J_{21}$ at a distance of $20 \rm \,pc$, the distance between the two gas clumps in our simulation.

Ultimately, the number of dissociating photons outweighs the production of new $\rm H_2$ and we find that the time it takes for the full sphere to be dissociated is $\sim$ 150 yrs. This analytical result does not fully take into account self-shielding or the progression of this front at different speeds (based on density or other local properties). However, it shows promise for the ability of LW photons to promptly dissociate $\rm H_2$ in the regions around forming stars/sink particles.

We then checked the time it would take for an ionizing front to sweep over this same spherical shell. This is useful to determine the treatment of ionizing feedback when creating star particles. We again use equation~\ref{eq:Mihir}, now assuming $n_\text{p} = n_\text{H} = n_\text{e}$ in the ionized region. The temperature is set to the atomic cooling limit, $\sim$$10^4 \,\rm K$. Ionizing photons have energies $ > 13.6 \, \rm eV$, but their number flux is comparable to the LW flux. $\dot N_{\rm ion}$ is therefore set to the previous $\dot N_{\rm LW}$ value. 



We find that the recombination rate outweighs the ionization rate and that this ionization front cannot propagate to the second clump. \cite{Kulkarni19} uses a more detailed analytical setup to estimate the propagation of the ionization front, and find it takes $\sim$ 2 Myr to travel 20 pc. Since this is beyond the simulated time period in our simulation, we feel comfortable not including UV radiation/feedback. Correspondingly, we do not artificially heat the halo to simulate this ionization heating.

\FloatBarrier
\section{Results}
\label{sec:Results}

In this section, we describe in detail the results of Table~\ref{tab:Test Runs}'s first three runs. We explain the fourth run's addition and its results in Sec~\ref{subsec:NoCooling}. We first briefly discuss fragmentation and its relevance to the subsequent results. We then analyze clump B, which forms a sink particle after clump A. This follows the goal of this work, examining how the introduction of an additional LW flux impacts sequential star formation. Clump B is the second high-density clump, located at the bottom of Fig.~\ref{ZoomOutDensTemp}. We initially look at the masses and accretion rates of the largest sink particle, ``sink B1'', within this clump. We also look at the gas properties that influence the sink particles' growth. This includes the gas temperature, gas number density (encompassing the nine chemical species listed in Sec.~\ref{sec:SimulationSetup}), $\rm H_2$ fraction ($X_{\text{H}_2}$), and electron fraction ($X_\text{e}$). Lastly, we briefly summarize the effect of the LW radiation on clump A. This is the high density clump at the center of Fig.~\ref{ZoomOutDensTemp} and it contains the first protostar that forms, sink A1. We can analyze the additional LW flux's impact on clump A in the same way as clump B due to us implementing a uniform background flux. Studying clump A demonstrates what would happen if it was the ``secondary'' clump subject to an additional LW flux from a hypothetical neighboring protostar. Even though there is no such previously-formed star in our simulation, the differences in gas density and temperature between clumps A and B provide a useful additional datapoint to study the LW radiation's effect on gas and star formation in different environments.

\subsection{Fragmentation}
\label{subsec: Fragmentation}
We observe fragmentation, defined by the number of sink particles that form, across our three initial runs. This fragmentation is largest in the background-only run, followed by the long-delay and finally the short-delay run. In the short-delay run, adding the LW flux to clump B before sink B1 forms eliminates any additional sink particle formation. This supports the expectation that the LW radiation reduces fragmentation and matches the result below that the internal LW flux inhibits cooling across the two clumps. The fragmentation does not affect our main results in any of these runs. The first sink particles to form in each clump, sink A1 and B1, remain the most massive sink particles across the simulated time. We therefore restrict our focus to these two sink particles.

\subsection{Clump B}
\label{sec:Clump B}

We investigate how adding an additional LW flux affects clump B and the protostar that forms within it. We first look at the protostar's growth and then its local gas properties such as temperature, density and $X_{\text{H}_2}$. We track these values beginning at our snapshot \#17, $7 \times 10^4$ years after the (re)start of our simulation (snapshot \#10). This is the first snapshot to contain a sink particle in clump B. It is also $t = 0.06 \rm \, Myr$ years after the first sink particle (sink A1) forms in clump A. The total number of snapshots varies between runs, the shortest being 74 (``background only'') and the longest being 110 (``no cooling'').

\subsubsection{Accretion rate and protostellar mass}
\label{sec:Bottom Mass}

\begin{figure}
     \centering
     \includegraphics[width=.48\textwidth]{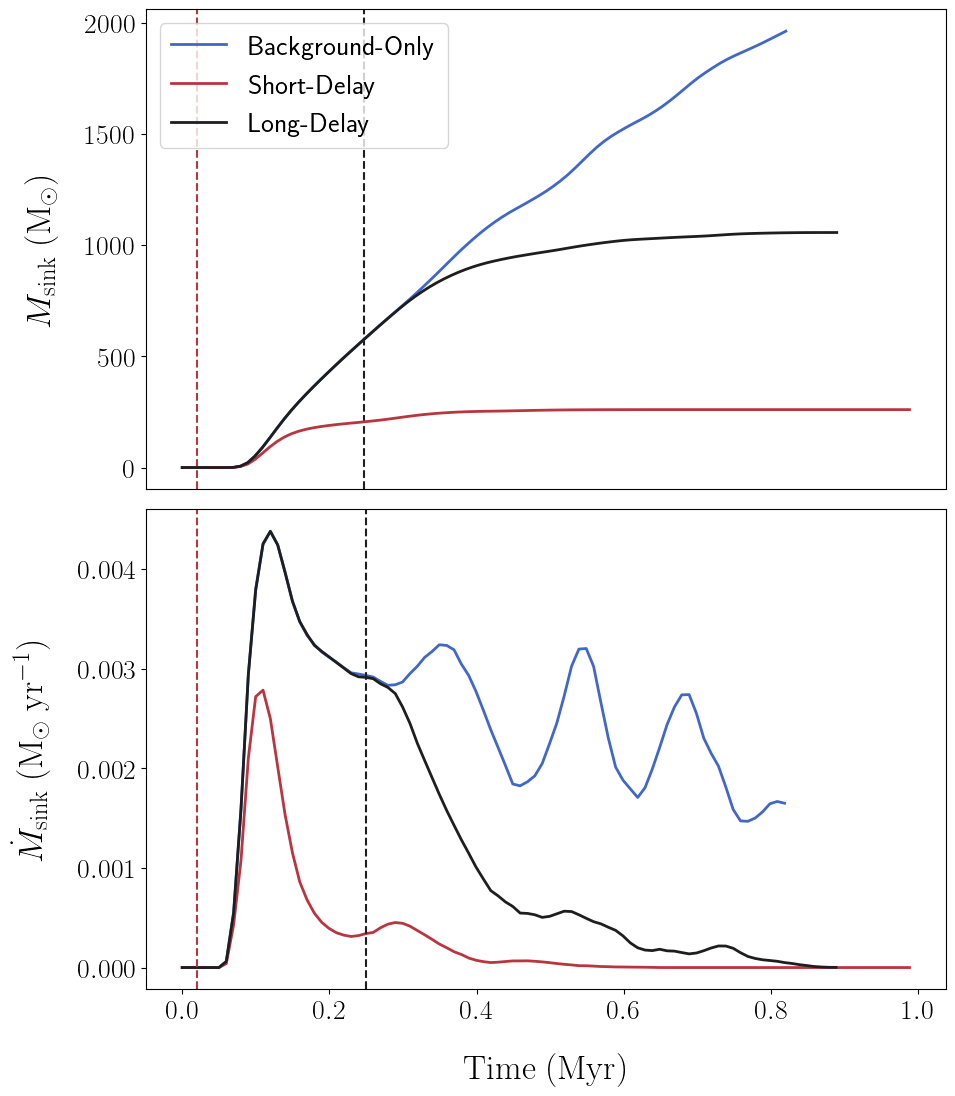}

    \caption{The top panel displays the sink B1's mass and the bottom panel displays its accretion rate as a function of time for each run. $t=0$ is defined as the (re)start of the simulation at $z = 6.5648$. The three solid colored lines represent the three test runs (blue: background only; red: short-delay; black: long-delay). The vertical red and black dashed lines represent when the additional internal LW flux is added in the short- and long-delay runs, respectively. When internal LW radiation is introduced, the accretion rate rapidly decreases and the masses reach significantly lower values than in the background-only run.} 
    \label{fig:BottomMassAccretion}
\end{figure}

We calculate and plot sink B1's mass and accretion rate versus time (Fig.~\ref{fig:BottomMassAccretion}) to measure how the additional LW flux affects the sink particle's growth. We do not observe the expected increase in accretion rate and sink particle mass when the additional flux is added - we instead find the opposite trend.

In the background-only run, sink B1's mass climbs to roughly $2000 \, \rm M_\odot$ within 0.84 Myr and is continuing to grow at the time we end the simulation (Fig.~\ref{fig:BottomMassAccretion}). However, the accretion rate peaks at $4.4 \times 10^{-3} \,\rm M_{\odot} \, yr^{-1}$ and then fluctuates between $\sim$$(2-3) \times 10^{-3} \,\rm M_{\odot} \, yr^{-1}$. This remains well under the minimum rate to delay collapse onto the main sequence, $0.01-0.04 \, \rm M_\odot \,yr^{-1}$. We therefore expect to see this protostar settle onto the main sequence at lower masses or fragment into several low mass stars. 

In the two other runs, the sink particle mass quickly flattens out after the additional LW flux is introduced. In the short-delay run, the LW flux decreases the observed accretion peak measured in the background-only run. The accretion rate then continues to decrease and effectively stops by 0.67 Myr (Fig.~\ref{fig:BottomMassAccretion}). The mass in the short-delay run therefore plateaus at $260 \,\rm M_\odot$, significantly lower than in the background-only run. We see that the additional LW flux produces a smaller protostar, rather than the desired more massive one. 

In the long-delay run, this trend is the same. However, the accretion rates and masses diverge after the initial peak in accretion rate due to the delay in the additional LW flux being introduced. Once the additional flux is added, the accretion rate quickly falls to $\lesssim 10^{-3} \,\rm M_{\odot} \,  yr^{-1}$ by 0.4 Myr. The rate continues to decrease until accretion stops, as in the short-delay run. The final mass value flattens out at $\sim$$1000 \,\rm M_\odot$. We see that delaying the additional flux does not increase the likelihood of a massive star forming. The increase in final mass compared to the short-delay run is simply due to delaying the decrease in the accretion rate.

\subsubsection{Density and temperature near the protostar}
\label{sec:Bottom Temp}

\begin{figure}
     \centering
     \includegraphics[width=.48\textwidth]{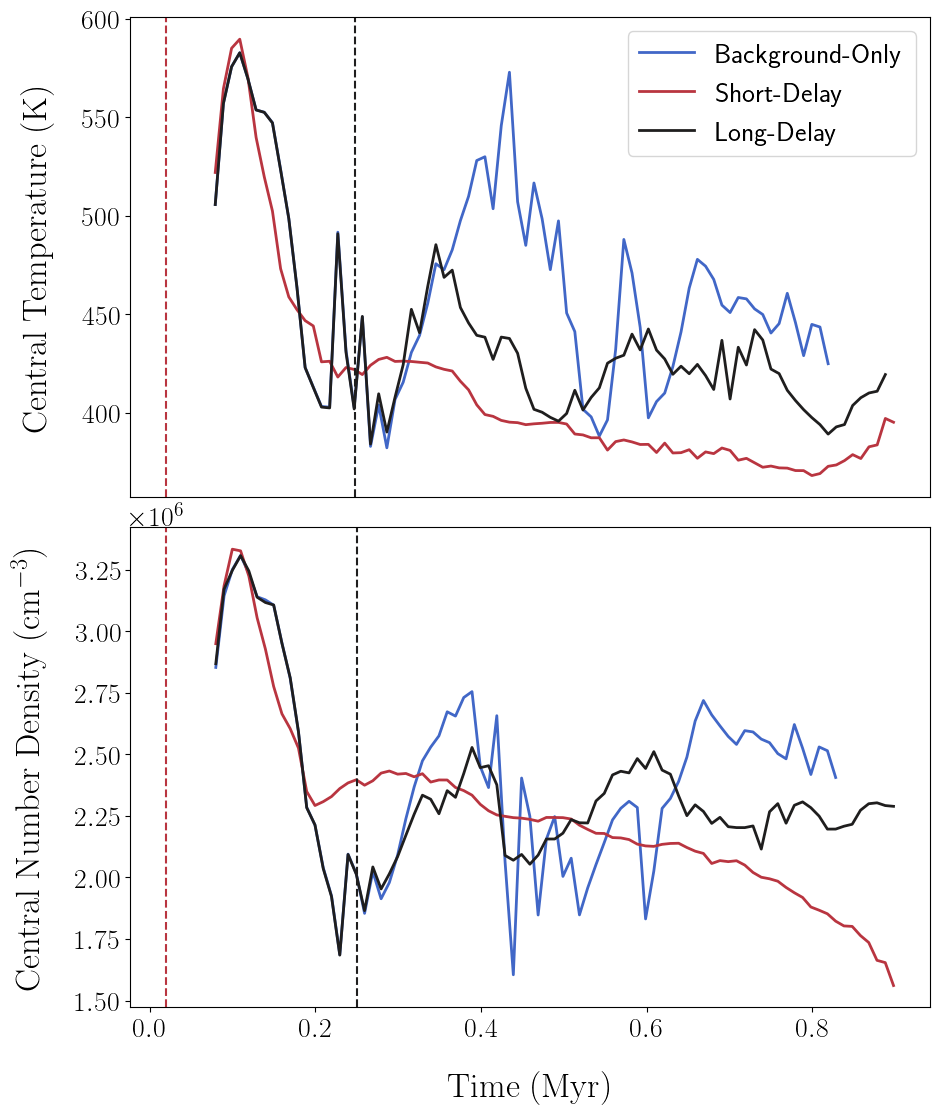}
     
    \caption{The top and bottom panels display the temperature and density evolution, respectively, at sink B1's location. The sink particle location is defined as the grid cell containing sink B1. The blue, red, and black lines represent the background-only, short-delay, and long-delay cases respectively, as in Fig.~\ref{fig:BottomMassAccretion}. The central temperature in the background-only run is higher than in the runs with additional LW flux, for several hundred thousand years.}
    \label{BottomPointTempDens}
\end{figure}

\begin{figure}
     \centering
     \includegraphics[width=.48\textwidth]{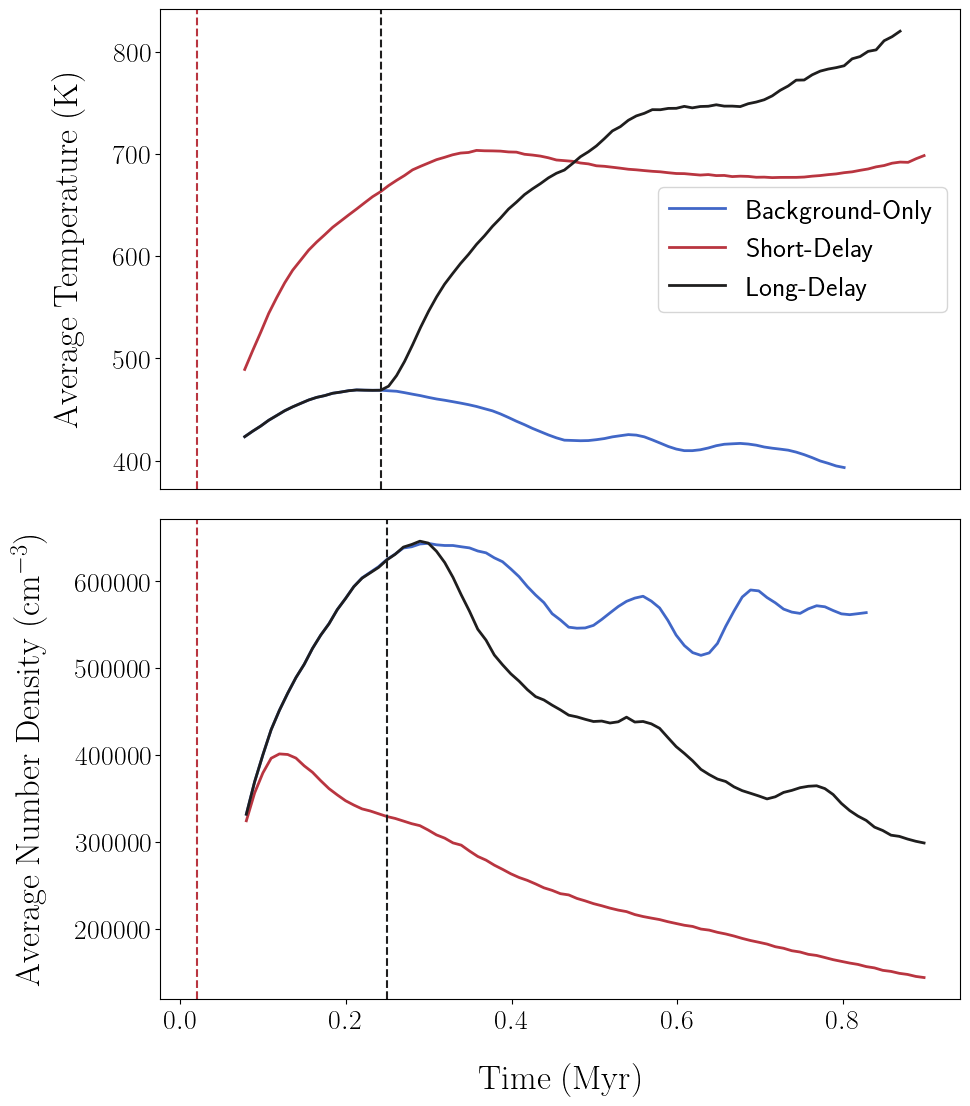}
     
    \caption{The top and bottom panels show the evolution of the average temperature and density within 1~pc of sink B1, respectively. The blue, red, and black lines represent the background-only, short-delay, and long-delay runs respectively, as in Figs.~\ref{fig:BottomMassAccretion} and~\ref{BottomPointTempDens}. The impact of the additional LW flux on the average temperatures and number densities is clearer than on the central values shown in Fig.~\ref{BottomPointTempDens}. The temperature again rises significantly immediately after additional LW is introduced. However, the short-delay temperature peaks at and then settles near $T =700 \, \rm K$ (Fig.~\ref{BottomAvgTempDens}). The temperature continues to climb in the long-delay run throughout the simulation time, reaching $T = 820 \, \rm K$. The densities decrease with the additional LW flux.}

    \label{BottomAvgTempDens}
\end{figure}

We measure the density and temperature of the gas cell at the sink particle's location to better understand why the additional LW flux reduces the accretion rate. We label this gas cell's values as the ``central'' values. The central values fluctuate significantly, partially due to the dynamics of the sink particle/clump system. The sink particle drifts through higher/lower temperature and density regions at times, producing some variation in the gas temperature, density, etc. at its location. We then also measure the average gas temperature and density in a 1~pc region centered on sink B1, to display how the LW flux affects a wider region. These ``average'' values also display clearer trends than the more volatile central ones.

The additional LW flux generally decreases the temperature at sink B1's location (Fig.~\ref{BottomPointTempDens}). This difference between the background-only and the two additional LW flux runs becomes the most apparent past $t \gtrsim 0.3 \, \rm Myr$. At $t = 0.74 \, \rm Myr$, the end of the background only run, the central temperatures in the short-delay and long-delay runs are $T = 373 \,\rm K$
and $ T = 389 \, \rm K$ respectively. These are both slightly below the background-only run's central temperature, $T = 424 \,\rm K$. Supermassive star formation requires significantly higher temperatures. The observed downwards trend in temperature produced by the additional LW radiation implies that the additional flux hurts the sink particle's ability to become a supermassive star. 

The average temperatures of the clumps experience different trends. The average temperature of the larger region in clump B shows the opposite trend, increasing significantly above the background-only run when the additional LW flux is added (Fig.~\ref{BottomAvgTempDens}). The average temperature in the background-only run heats up to $469 \, \rm K$ and then slowly decreases past $400 \, \rm K$ by $t = 0.84 \, \rm Myr$. The short-delay produces an average temperature that settles at $ T \simeq 700 K$. The average temperature in the long-delay is at $820 \, \rm K$ and continuing to rise at the time the run ends ($t = 0.90 \, \rm Myr$). We therefore see that the additional LW flux increases the average temperature of the nearby gas, albeit not at the sink particle. 

The additional LW flux does not produce a clear increase or decrease in the central density until late in the simulation (Fig.~\ref{BottomPointTempDens}). Past $t \gtrsim 0.65 \, \rm Myr$, the additional LW flux decreases the central density consistently below the background-only run. Prior to this point, the densities in the three runs remain relatively close together and alternate being higher or lower. However, there is a general trend maintained by all three runs. The density initially spikes at $\sim$$0.1 \, \rm Myr$ before trending downwards at later times. The overall decrease in density again hurts the protostar's ability to accrete gas and form a supermassive star.

The trend in the average density is also clearer than at the sink location. The average densities decrease with the additional LW flux (Fig.~\ref{BottomAvgTempDens}). Furthermore, the decrease in density after the additional LW flux is added persists throughout the simulation.

\subsubsection{Effects on clump B's gas morphology }
\label{sec:Bottom 1D Profiles}

\begin{figure*}
    \centering
    \includegraphics[width=.95\textwidth]{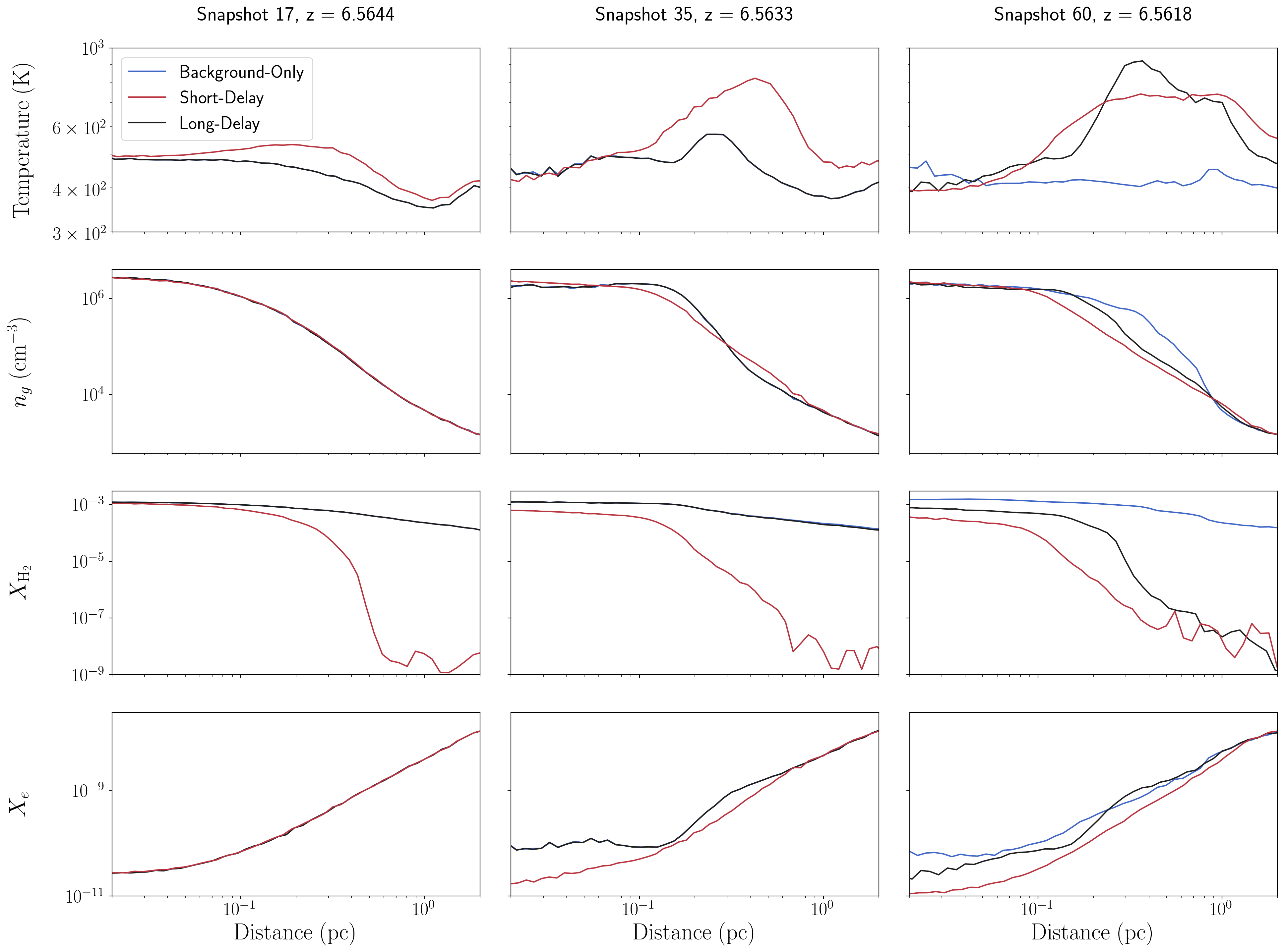}

    \caption{From top to bottom: radial profiles of temperature, gas number density, $X_{\text{H}_2}$, and $X_\text{e}$ centered about sink B1. 
    From left to right, the columns display the profiles at snapshot \#17 ($z$ = 6.5644, $\sim$$0.06 \rm \, Myr$ after sink A1 forms in snapshot \#11), 0.18~Myr later at snapshot \#35 ($z$ = 6.5633), and an additional 0.25 Myr later at snapshot \#60 ($z$ = 6.5618). The blue, red, and black lines represent the background-only, short-delay, and long-delay cases respectively. Snapshot \#35 (middle column) displays the effects of the additional LW flux in the short-delay run just before the time when the additional LW flux is added in the long-delay run. We see the influence of the additional LW flux in both the short-delay and long-delay runs in snapshot \#60 (right column). We see increases in temperature in the short and long-delay runs. However, the clump’s innermost region retains a low temperature. The background-only run maintains the highest temperature at the clump center. We see different shifts in gas number density after the internal LW flux is added. The density is slightly raised above the background-only run at the center and outermost regions but decreases more significantly over the majority of the clump's volume. $X_{\text{H}_2}$ also decreases after the additional flux is added. However, this decrease is highly sensitive to the gas density. It remains high close to sink B1, where the gas density is highest, and drops off rapidly at the lower densities further from the clump center. The short-delay run has a reduced electron fraction relative to the background-only run until roughly 1~pc. In the long-delay run, $X_\text{e}$, is only reduced relative to background-only run at low $r$.}
    \label{fig:BottomPhase}
\end{figure*}

Finally, we provide 1D profiles that display some of the changes in temperature, density, $X_{\text{H}_2}$, and $X_\text{e}$ over a 2~pc region (Fig.~\ref{fig:BottomPhase}). These profiles are centered on sink B1 and are spherically averaged. We choose a 2 pc radius because this covers a large region of the wider clumps A and B. This is evidenced by the large range of gas number densities covered by the resulting spherical profiles (Fig.~\ref{fig:BottomPhase} as well as Fig.~\ref{fig:TopPhase} for clump A; see below). We show the initial profiles at snapshot \#17, $z=6.5644$. We then display the point we turn on the additional LW in the long-delay case, $t = 0.25 \, \rm Myr$ after the simulation (re)start. By this point, we see a difference in the short-delay run due to the additional LW. Finally, we show the distributions an additional 0.25 Myr later, by which point both the short- and long-delay runs display differences.

The radial profiles for temperature and density reinforce what was shown in Figs.~\ref{BottomPointTempDens} and~\ref{BottomAvgTempDens}. Close to the sink particle, the additional LW flux in the short-delay run raises the gas temperature slightly higher than in the background-only run (Row~1, Fig.~\ref{fig:BottomPhase}). We see by the second column, at snapshot \#35, the central temperature is lower for the short-delay run. However, this flips as the distance from the sink particle increases. The LW flux raises the gas temperature of the majority of the spherical region above that of the background-only run. This is again seen in snapshot \#60 for both of the additional LW flux runs. The LW flux is only able to raise the gas temperature outside of the clump's center.

The density profile also shows that the LW flux impacts the clump differently further away from the sink particle (Row~2, Fig.~\ref{fig:BottomPhase}). Looking at snapshot \#35, the additional LW flux raises the density above the background-only run's at the center and in a small region further away. However, there is a region in between where the density is decreased by the additional LW flux. We see that the decrease in density here is large enough to produce the decrease in our clump's average density (Fig.~\ref{BottomAvgTempDens}). This becomes more clear in snapshot \#60, which shows a large decrease in gas density further away from the sink particle when the additional LW flux is present. 

We next show $X_{\text{H}_2}$ to check whether the LW flux successfully dissociates $\text{H}_2$ (Row~3, Fig.~\ref{fig:BottomPhase}). The fraction changes drastically as the distance from the sink particle increases. At snapshot \#35 in the short-delay run, $X_{\text{H}_2}$ decreases from $\sim$$1 \times 10^{-3}$ to $\sim$$2 \times 10^{-4}$ by 0.25~pc and is roughly 5 orders of magnitude lower by 0.5~pc. By comparison, the fraction in the background-only case is $\sim$$4 \times 10^{-4}$ at 0.5~pc and greater than $\sim$$6 \times 10^{-4}$ at 0.25~pc. However, the central $X_{\text{H}_2}$ remains high for the short-delay run. This is also apparent in the long-delay run at snapshot \#60. $X_{\text{H}_2}$ drops off sharply past $\sim$$0.2 \, \rm pc$, eventually reaching fractions on the order of $10^{-8}$. $X_{\text{H}_2}$ in the background-only case remains above $10^{-4}$ out to the edge of our 2~pc sphere. We clearly see that at large distances, the LW flux is effective in dissociating $\rm H_2$ down to low fractions $\lesssim 10^{-8}$ (Fig.~\ref{fig:BottomPhase}). However, the additional LW flux does not efficiently dissociate $\rm H_2$ close to the sink particle. The central fraction in snapshot \#35 is only a factor of $\sim$two lower in the short-delay run. At snapshot \#50, the central $X_{\text{H}_2}$ in the short-delay run has further decreased to $\sim$$5 \times 10^{-4}$. The central fraction is $\sim$$1 \times 10^{-3}$ for the long-delay case, as compared to $\sim$$1.4 \times 10^{-3}$. The gas still cools through molecular cooling at these high $X_{\text{H}_2}$ values. We have thus simply slowed the cooling process rather than halting it at the sink particle's location. 

Finally, the electron fraction, $X_\text{e}$, decreases near the sink particle after the LW flux is added (Row~4, Fig.~\ref{fig:BottomPhase}). In snapshot \#35, this decrease in the short-delay case extends to $\sim$$0.8\, \rm pc$. By snapshot \#50, the decrease is visible in both the additional LW runs. $X_\text{e}$ is lowered relative to the background-only run past $\sim$$1\,\rm pc$ for the short-delay run and out until $\sim$$0.2 \, \rm pc$ for the long-delay run. 

We further illustrate the internal LW flux's affect across clump B in Figs.~\ref{fig:NoLWBottom}-\ref{fig:250DelayBottom} of Appendix~\ref{sec:Appendix}. The figures display a projection of density, slice of temperature, and slice of $X_{\text{H}_2}$ centered on sink B1. It displays these for snapshots \#17 (top row) and \#60 (bottom row) in the initial three runs.

\subsection{Clump A}
\label{sec:Top Clump}

We now briefly describe the additional LW background's analogous effects on clump A and its protostar. Its most massive sink particle, sink A1, forms in snapshot \#11. This is in the middle of the additional LW flux increase in the short-delay run and prior to the additional LW flux being added in the long-delay run. We do not expect a massive star to form here, since the initial delay in accretion due to heating the gas will cause the existing protostar to collapse onto the main sequence at lower masses. However, we treat it as a second test site for studying how LW radiation affects the gas chemistry and cooling. We first discuss the LW background's effects on stellar growth and then the effects on nearby gas properties such as density, temperature, and $X_{\text{H}_2}$. 

\begin{figure}
    \centering
    \includegraphics[width=.48\textwidth]{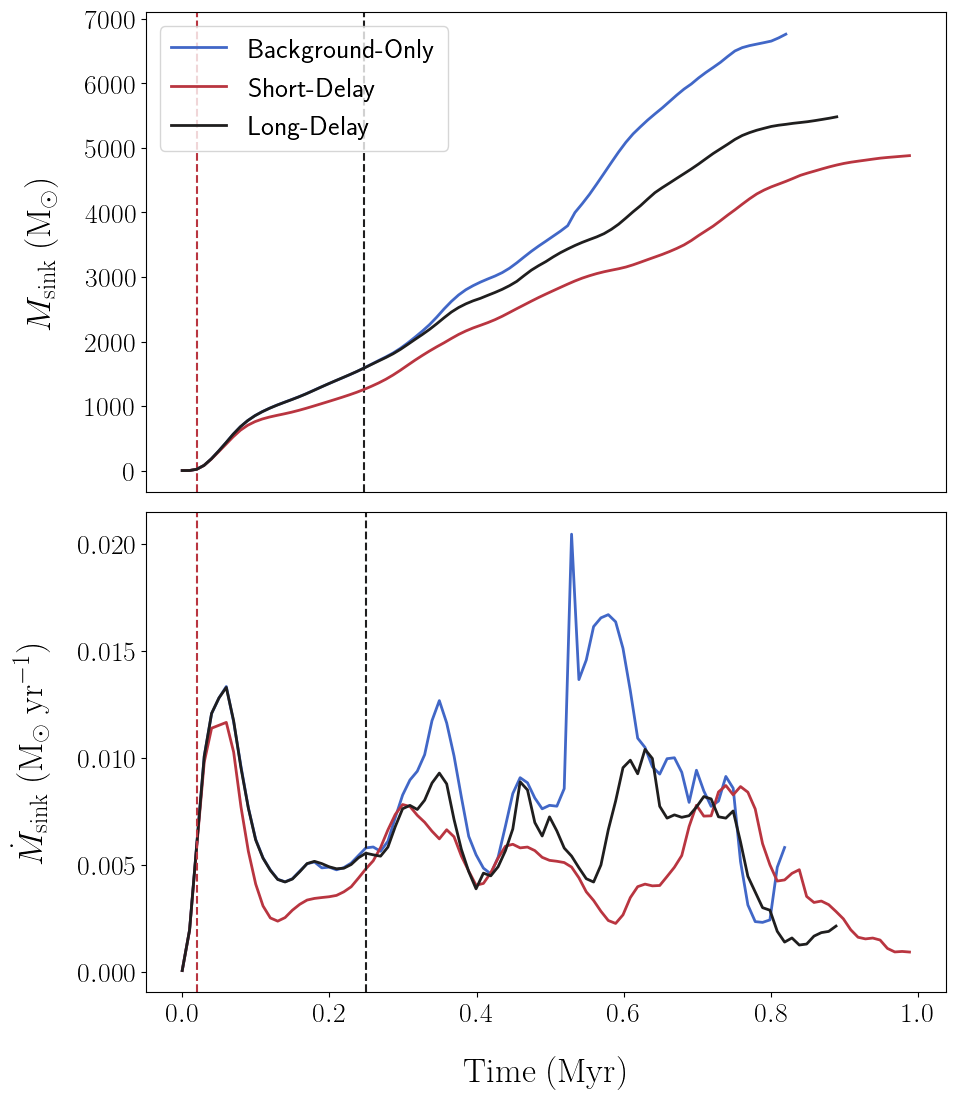}
    \caption{The evolution of the sink A1's accretion rate (bottom panel) and its mass (top panel) in the first three runs. The time is again defined as in Figs.~\ref{fig:BottomMassAccretion}-\ref{BottomAvgTempDens}. The three colors represent the three test runs with the same color scheme as in Figs.~\ref{fig:BottomMassAccretion}-\ref{fig:BottomPhase}. The additional internal LW flux decreases the sink's accretion rate and final mass in the short- and long-delay runs.}
    \label{fig:TopMassAccretion}
\end{figure}

\begin{figure*}
    \centering
    \includegraphics[width=.95\textwidth]{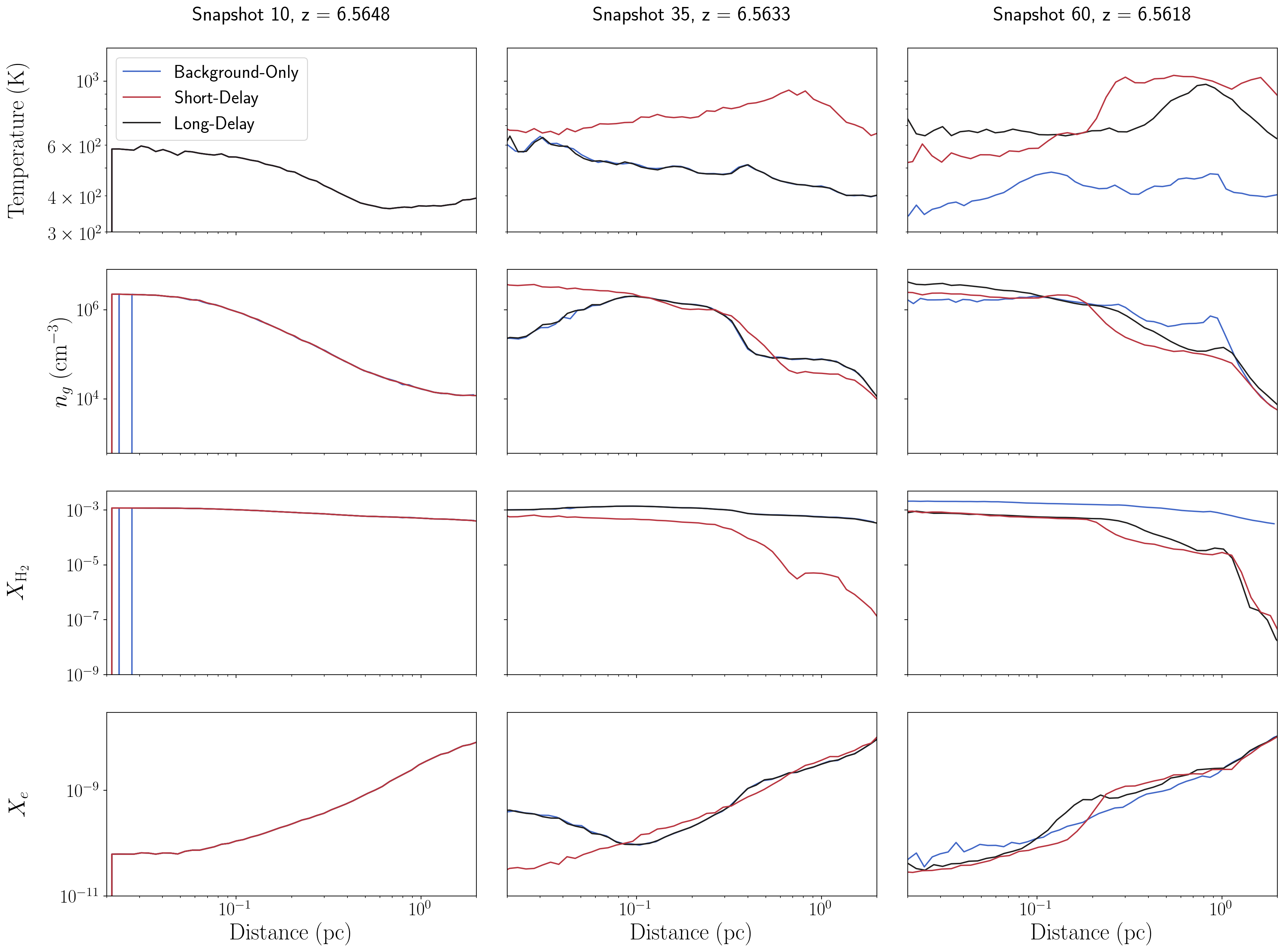}

    \caption{From top to bottom: radial profiles of temperature, gas number density, $X_{\text{H}_2}$, and electron density centered on sink A1. We display the region from 0.02 to 2~pc around sink A1. This limits the number of empty cells that the produce vertical spikes seen in the leftmost column. From left to right, the panels display the profiles at snapshot \#10 ($z = 6.5648$), 0.25~Myr later at snapshot \#35 ($z = 6.5633$), and an additional 0.25~Myr later at snapshot \#60 ($z = 6.5618$). The blue, red, and black lines again represent the background-only, short-delay, and long-delay cases respectively. We see similar trends to the bottom clump (Fig.~\ref{fig:BottomPhase}). The additional LW drives an increase in temperature over the entire 2 pc region. It also drives a small region of increased density near sink A1. However, the density decreases outside of this $\sim$$0.1 \rm \,pc$ core for the two runs with extra LW compared to the background-only run. $X_{\text{H}_2}$ remains high in the high-density region around the sink particles before falling off at larger distances. The trend in $X_\text{e}$ roughly mirrors that of the gas density. $X_\text{e}$ decreases close to the sink particle but is raised in portions of the clump at larger $r$.}
    \label{fig:TopPhase}
\end{figure*}

As expected from clump B, introducing the LW radiation reduces the accretion rates and final mass of sink A1 (Fig.~\ref{fig:TopMassAccretion}). This drop-off in both quantities is less drastic than sink B1's, whose accretion rate drops to zero. However, a notable difference is that clump A begins its collapse and forms a protostar prior to the LW background being added. It also has a higher gas density than clump B (see Figs.~\ref{fig:BottomPhase},~\ref{fig:TopPhase}). This prevents sink A1's accretion rate from dropping all the way to zero. However, the same trends of decreases in the accretion rate and the final sink mass are observed. The additional LW flux prevents the protostar from becoming a massive star.

\begin{figure}
     \centering
     \includegraphics[width=.48\textwidth]{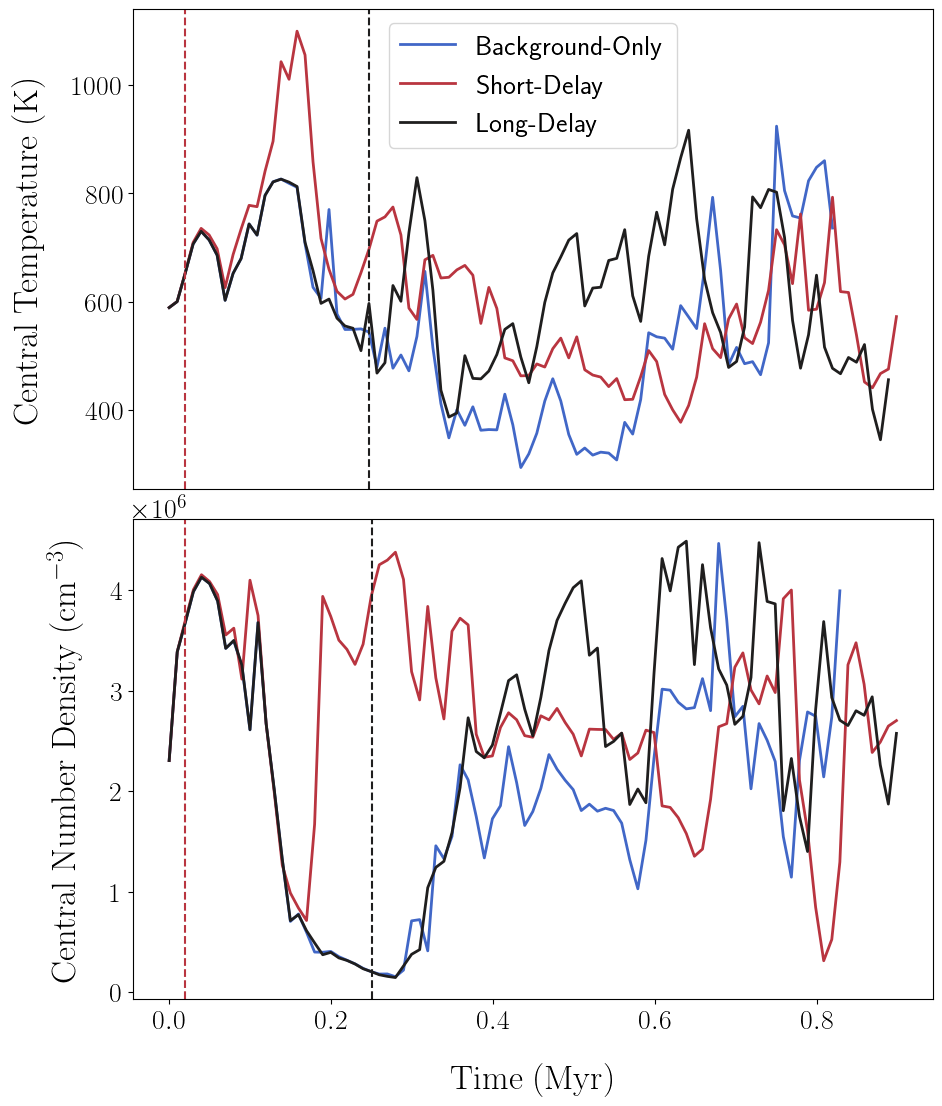}
     
     \caption{Evolution of the temperature (top panel) and density (bottom panel) at sink A1's location. The sink particle location is defined as the grid cell containing sink A1. Again the blue, red, and black lines represent the background only, short-delay, and long-delay cases respectively, as in Figs.~\ref{fig:BottomMassAccretion}-\ref{fig:TopMassAccretion}. There is large fluctuation across these values. However, the additional LW flux in the short- and long-delay runs produces an initial temperature and density spike.}
     \label{TopPointTempDens}
\end{figure}

\begin{figure}
     \centering
     \includegraphics[width=.48\textwidth]{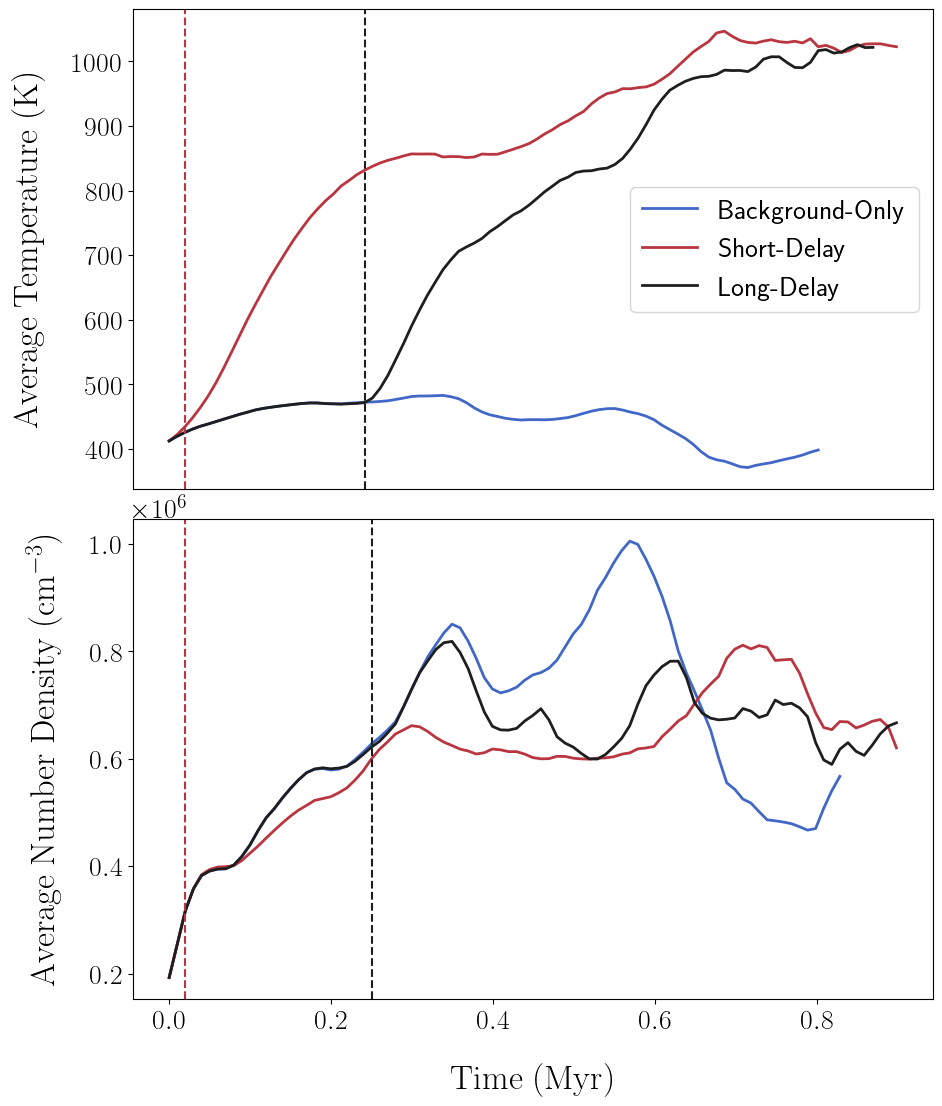}

    \caption{The top and bottom panels show the time evolution of the average temperature and density, respectively, within 1 pc of sink A1. Again the blue, red, and black lines represent the background only, short-delay, and long-delay cases respectively. The changes in these average temperatures and number densities are more obvious than those for clump B in Fig.~\ref{TopPointTempDens}. The average temperatures rise significantly immediately after the additional LW is introduced, and both hover near 1020 K by $\sim$$1 \, \rm Myr$. The average number densities initially decrease with the additional LW flux but later surpass the background-only case at $\sim$ 0.65~Myr.} 
    \label{TopAvgTempDens}
\end{figure}

The temperature and density at the sink particle location again fluctuate significantly. We see clear spikes in the temperature and density when the background flux is added (Fig.~\ref{TopPointTempDens}). However, the average temperature of the clump (measured again within 1~pc of the sink) clearly increases while the density decreases (Fig.~\ref{TopAvgTempDens}). The spikes in temperature match the expectation that the LW flux will reduce the gas's ability to cool through $\rm H_2$. However, the temperature stalls well below the atomic cooling threshold of $\sim$$ 10^4 \, \rm K$, preventing the gas from experiencing the desired isothermal collapse at higher temperatures. We therefore just increase the thermal energy of the gas, slowing accretion onto the protostar. 

We again show profiles of the spherically-averaged temperature, density, $X_{\text{H}_2}$, and  $X_\text{e}$ surrounding sink A1 at three different times (Fig.~\ref{fig:TopPhase}). The 1D profiles at snapshot \#17 are replaced by snapshot \#10 (the start of our simulation). The snapshot \#10 profile is centered on clump A's maximum gas density, where sink A1 forms by the next snapshot. 

We see the same general trends as in clump B with a couple of notable differences. The additional LW flux drives the temperature upwards more clearly, increasing it by nearly a factor of two at certain radii (Fig.~\ref{fig:TopPhase}). The density increases at the core of the clump when LW is added. However, this does not hold across the entire region. The background-only run contains higher density gas at certain radii. The $X_\text{e}$ trend again roughly mirrors the gas number density and is similar to clump B's $X_\text{e}$ profile.

The $\rm H_2$ profiles again demonstrate the key reason why we do not see an increase in the accretion rates, while also showing the effect of a higher gas density. $X_{\text{H}_2}$ drops off at larger radii when the LW turns on while the central region maintains its high $X_{\text{H}_2}$ (Fig.~\ref{fig:TopPhase}). These radii where we see the drop-off are pushed out farther in comparison to clump B. This can be attributed to the higher density in this clump, which therefore retains a larger $\rm H_2$ core. The LW fluxes we use are clearly unable to dissociate the hydrogen at the high densities we reach at the centers of both clumps. 

We further diagnose the internal LW flux's effect on clump A in Figs.~\ref{fig:NoLWTop}-\ref{fig:250DelayTop} of Appendix~\ref{sec:Appendix}. These figures again display a projection of clump density, slice of temperature, and slice of $X_{\text{H}_2}$, now centered on clump A/sink A1. It displays these for snapshots \#10 (top row) and \#60 (bottom row) in the initial three runs.

\section{Discussion}
\label{sec:Discussion}

Contrary to our initial expectations, we observed a negative impact on the accretion rates and sink particle masses when we add the ``internal'' LW radiation. In clump B, the background-only run produces a $2000 \, \rm M_{\odot}$ sink particle by 0.83~Myr. By this time, the short- and long-delay runs produce sink particles with mass $250 \, \rm M_{\odot}$ and $1000 \, \rm M_{\odot}$, respectively. These lower masses correspond to the lower accretion rates displayed in Fig.~\ref{BottomPointTempDens}. We similarly compare the changes in gas temperature and density at the sink particle location and within the larger protostellar core. We see that the protostellar core's temperature rises above several hundred K but does not surpass $1000\, \rm K$. The temperature at the sink particle location itself decreases. This is evidence that within the 1~Myr elapsed in our simulations, the gas cannot heat to the atomic cooling threshold and then rapidly collapse via atomic cooling into a more massive protostar. The $X_{\text{H}_2}$ profile further confirms this negative conclusion. We find significantly dissociated $\rm H_2$ only at low densities, further away from the sink particles. The dense protostellar core maintains a relatively high $X_{\text{H}_2}$ at its center and continues to cool via ${\rm H_2}$. We are thus left with a core of dense and cold gas accreting onto the protostar. However, the surrounding gas cools less efficiently, slowing accretion onto the protostar. The additional ``internal'' LW flux thus decreases the final protostellar masses.

Clump A shows very similar trends, albeit with minor differences. These differences are due to the different environment (e.g. temperature, gas density) and timing of the ``internal'' LW flux being added after sink A1's formation. The clump has a higher initial gas density at the time the additional LW radiation is turned on and $X_{\text{H}_2}$ consequently remains relatively high out to larger distances away from the sink. The internal LW flux raises the gas density at the sink particle location but produces a decrease at larger radii. The gas temperature increases across the whole $\sim$$1 \rm \, pc$ region, with the highest temperatures at large $r$. The reason for the reduced accretion rate mirrors clump B. The $\rm H_2$ and therefore molecular cooling is not removed at the sink particle location. Instead, the reduced $X_{\text{H}_2}$ makes cooling less efficient in the lower-density regions further away, which provide the accretion into the region closer to the sink. This gas cools less and therefore accretes less efficiently. The surrounding gas may warm up to the atomic-cooling limit, and then cool via HI and rapidly fall onto the sink particle later on. However, this would occur well after the 1~Myr we simulated, and well after the protostar represented by the sink would contract onto the main sequence.

The gas morphology in the lower density regions at large $r$ tells another important story. Most of the $\rm H_2$ in these regions has been dissociated (see Figs.~\ref{fig:BottomPhase} and~\ref{fig:TopPhase}, \ref{fig:NoLWTop}-\ref{fig:250DelayBottom}). However, the gas temperature in these regions is still well below the atomic cooling limit ($\sim$$10^4 \, \rm K$). This suggests that the dynamical time within the clump may be too large for us to see elevated infall and accretion rates within our simulation time, even if we fully dissociate $\rm H_2$ around the sink particles. 

Ultimately, we find that $\rm H_2$ is dissociated and the gas temperature is raised only in lower density regions. We can estimate the ``$\rm H_2$ survival density'', the gas density at which the $\rm H_2$ will not be dissociated, by balancing the $\rm H_2$ dissociation and formation rates from equation~\ref{eq:dissociation}. We use rate $k_{28}$ from \cite{Shang10} for the $\rm H_2$ dissociation rate. This creates the following relationship between the dissociation rate
and the recombination rate per $\rm H_2$ molecule: $10^{-12}\beta\, \text{J}_{21} =  \, k_9 {\frac{{n_\text{H}} {n_\text{e}}}{{n_{\text{H}_2}}}}$, where $\beta$ = 0.9. We then select values for temperature (used in $k_9$), $n_\text{e}$, and ${n_{\text{H}_2}}$. For this estimate, we use values from snapshot \#35 in the short-delay run. We select values at a radius of $\sim$ 0.25 pc, where we observe the $X_{\text{H}_2}$ begin to drop off (Fig.~\ref{fig:BottomPhase}). We define the $X_{\text{H}_2}$ cutoff as $10^{-5}$. Using ${n_\text{e}} \cong  0.2 \, \rm cm^{-3}$ and $T \cong 800 \rm \, K$ produces a hydrogen number density ${n_\text{H}} \cong 1.4 \times 10^5 \, \rm cm^{-3}$. The densities at the center of our clumps are significantly higher. We would thus need a much higher LW flux to dissociate the $\rm H_2$ in these high density, central regions.

\begin{figure*}
    \centering
    \includegraphics[width=.95\textwidth]{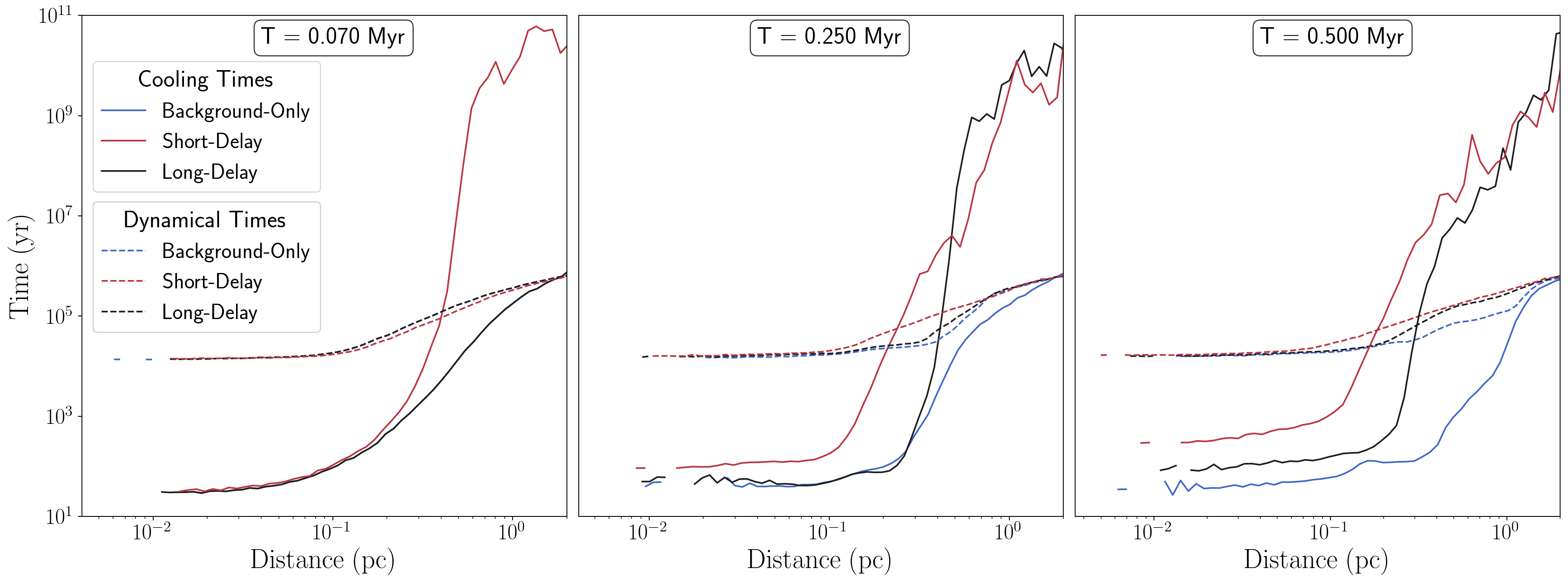}

    \caption{From left to right, the three panels show radial profiles of the cooling time ($t_{\rm H_2}$) and dynamical time ($t_\text{dyn}$) at snapshot \#17 ($z = 6.5644$), 0.18~Myr later at snapshot \#35 ($z = 6.5633$), and an additional 0.25~Myr later at snapshot \#60 ($z = 6.5618$). The blue, red, and black lines represent the background-only, short-delay, and long-delay runs respectively. The solid lines represent the cooling times, while the dotted lines represent dynamical times. The LW flux clearly increases the cooling time of the gas surrounding sink B1. This produces cooling times longer than the dynamical time slightly outside of the sink particles position. This point of intersection gets pushed further in with time as the gas continues to heat.}
    \label{fig:BottomCoolDyn}
\end{figure*}

\subsection{Cooling and dynamical times}

We next calculate the cooling and dynamical times to further investigate the internal LW radiation's effect on clump B. We compute the cooling time as
\begin{equation}
   t_{\text{H}_2} = \frac{1.5\,n_g\,k_B\,T}{\Lambda \,n_\text{H}\,n_{\text{H}_2}},
    \label{eq:Fernandez}   
\end{equation}
where $n_g$ represents the total gas number density. The cooling rate is defined using 
\cite{Galli98}'s analytical expression for gas in the low density limit:
\begin{equation}
   \begin{aligned}
   \text{log} \,\Lambda_{\text{H}_2} = -103.0 \,+\,97.59\,\text{log}\,T\,-\,48.05(\text{log}\,T)^2
   \\
   \,+\,10.80(\text{log}\,T)^3\,-\,0.9032(\text{log}\,T)^4,
    \label{eq:Galli}
    \end{aligned}
\end{equation}
where T is the gas temperature in Kelvin and $10\,\text{K} \,\leq \,T\,\leq \,10^4\,\text{K}$.
The dynamical time can be estimated as $t_{\text{dyn}} = \frac{1}{\sqrt{4\pi G\rho}}$, where $G$ is the gravitational constant, and $\rho$ is the gas density.

The internal LW flux raises the cooling time above the dynamical time in the gas surrounding the sink particle (Fig.~\ref{fig:BottomCoolDyn}). This confirms our earlier analysis that the internal LW flux inhibits the gas's ability to cool and accrete onto the protostar. Our protostar should join the main sequence before it reaches masses on the order of $10^3 \rm \, M_\odot$ or above due to the resultant low accretion. The cooling time surpasses the dynamical time further away from the sink particle, where the gas density and $X_{\text{H}_2}$ decrease. This implies that in these lower density regions, there may be time for the gas to heat to higher temperatures by compression as it begins to collapse, but one must then consider the UV radiation from the central massive star.

\subsection{$\textbf{H}_2$ cooling removed}
\label{subsec:NoCooling}
 
Motivated by the above findings, in which $\rm H_2$ molecules survive in the dense vicinity of the sink particles, we add a final run, named ``no-cooling''. With this run, we can conclusively determine whether massive star formation is possible in our halo when $\rm H_2$ cooling is removed. For simplicity, we add a constant $10^{10} \, \rm J_{21}$ internal LW flux beginning at dynamic collapse (z = 6.5618). This reduces $X_{\text{H}_2}$ below $ 10^{-13}$, shutting off molecular cooling entirely throughout the whole simulated volume. We determine whether or not the gas can heat to the atomic cooling limit, which is necessary to support the formation of a SMS from warm, atomic hydrogen. Sink A1 is able to form because we add this additional flux just prior to its formation. However, we shut off cooling before sink B1 is able to form. We therefore focus our analysis on clump A and the only sink particle to form, sink A1.

\begin{figure}
    \centering
    \includegraphics[width=.48\textwidth]{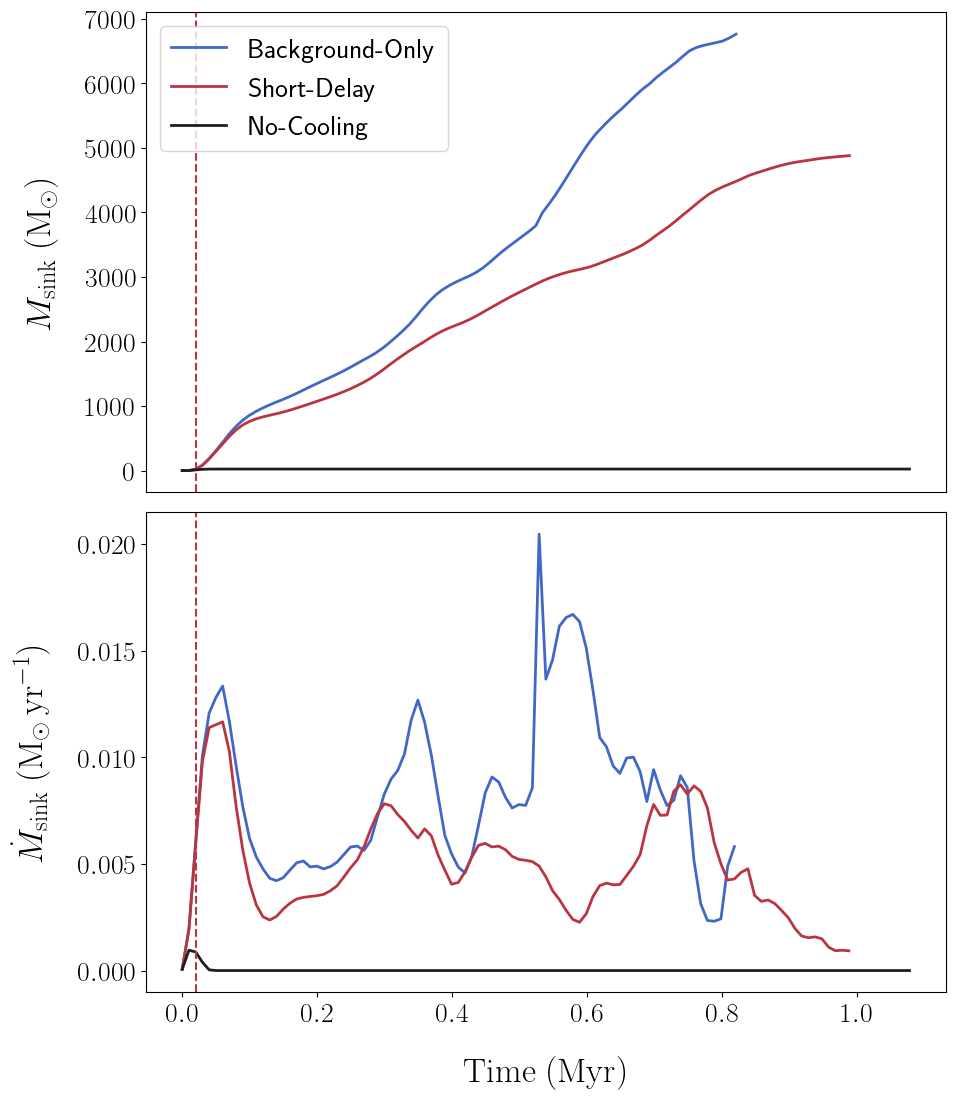}

    \caption{The evolution of sink A1's accretion rate (bottom panel) and mass (top panel) in the background-only (blue), short-delay (red), and no-cooling (black) runs. In the no-cooling run, the accretion shuts off soon after the flux is added and the mass stalls at $23 \,\rm M_{\odot}$.}
    \label{fig:1e10MassAcc}
\end{figure}

Shutting off cooling drastically impacts sink A1's evolution. The accretion rate peaks at $9.53 \times 10^{-4} \, \rm M_{\odot} \, yr^{-1}$ and then drops to zero by $0.05 \, \rm Myr$ (Fig.~\ref{fig:1e10MassAcc}). This yields a final mass of $23.1 \, \rm M_{\odot}$. This reduced sink mass and the absence of additional sink particle formation clearly demonstrates that the high LW flux has suppressed both cooling and further collapse within the two protostellar cores.

\begin{figure}
    \centering
    \includegraphics[width=.48\textwidth]{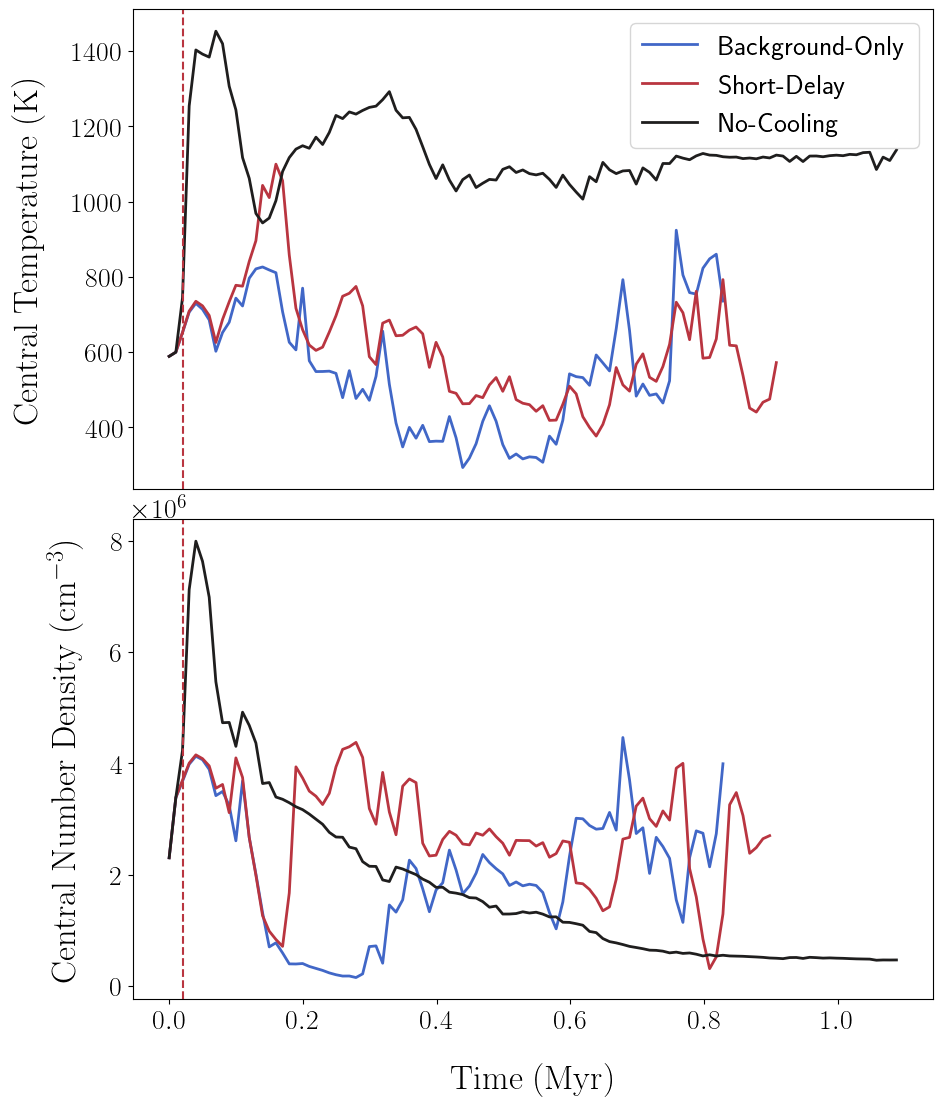}

    \caption{The top and bottom panels show the evolution of the temperature and density of the grid cell containing sink A1 respectively. The blue, red, and black lines represent the background-only, short-delay, and no-cooling runs respectively, as in Fig.~\ref{fig:1e10MassAcc}. The temperature is highest in the no-cooling run, matching expectations. This is due to $\rm H_2$ cooling being shut off. The density in the no-cooling run also spikes initially but then drops below the other two runs.}
    \label{fig:1e10PointTempDens}
\end{figure}

\begin{figure}
    \centering
    \includegraphics[width=.48\textwidth]{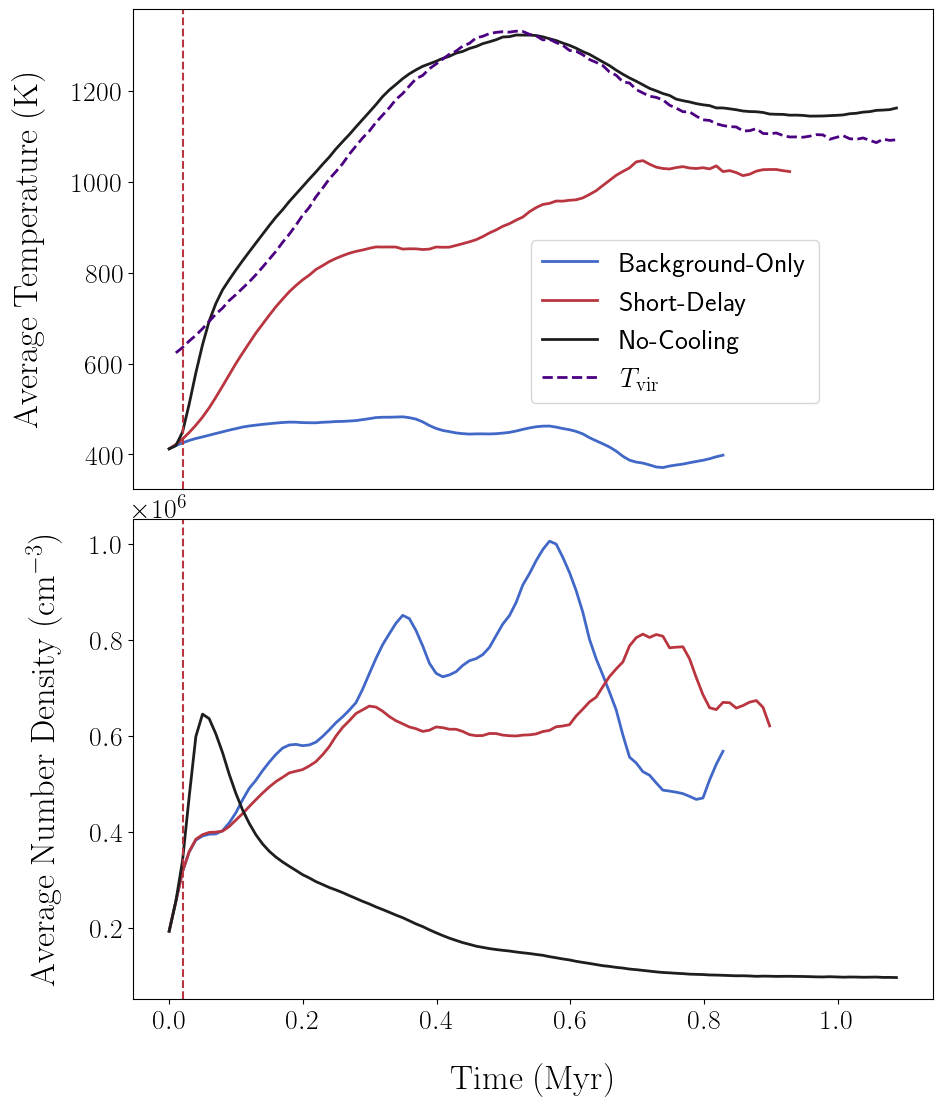}

    \caption{The top and bottom panels show the time evolution of the average temperature and density respectively within 1~pc of sink A1. Again the blue, red, and black lines represent the background-only, short-delay, and no-cooling runs respectively, as in Figs.~\ref{fig:1e10MassAcc} and~\ref{fig:1e10PointTempDens}. The gas virial temperature is marked by a dashed purple line. The high flux clearly produces a larger temperature increase in the gas surrounding the sink particle. However, the temperature does not reach the atomic cooling threshold at $10^4$ K. It settles near the virial temperature, which flattens out at $\sim$$1200\, \rm K$. Coupled with a large decrease in the gas density (after an initial spike), this shuts off accretion onto the sink particle (Fig.~\ref{fig:1e10MassAcc}).}
    \label{fig:1e10AvgTempDens}
\end{figure}

We again plot the temperature and density at the sink particle location (Fig.~\ref{fig:1e10PointTempDens}). The temperature in the high flux run is raised above that of both the background-only and the two $10^4 \, \rm J_{21}$ runs. It reaches a peak temperature of $1500 \, \rm K$ and then flattens out at $1100 \, \rm K$. However, this is still well below the atomic cooling threshold. The gas density at the sink particle initially increases to $1.63 \times 10^{-17}\, \rm g \, cm^{-3}$. This corresponds with the initial temperature spike, implying that the gas heats due to compression. However, the density then rapidly falls off as the thermal pressure balances out this compression. These same trends occur in the wider region of gas surrounding the sink particle (Fig.~\ref{fig:1e10AvgTempDens}). We measure the average gas temperature and density within a sphere of radius $1 \, \rm pc$ centered on the sink particle. The average temperature increases to $1300 \, \rm K$ and then flattens out closer to $1200 \, \rm K$. These values are both higher than in any of the previous runs. The average gas density also initially spikes and then falls well below the corresponding gas densities in the background-only and short-delay runs. The higher LW flux produces a region of high temperature, more diffuse pressure-supported gas that no longer cools and collapses onto the sink particle. 

\begin{figure*}
    \centering
    \includegraphics[width=0.95\textwidth]{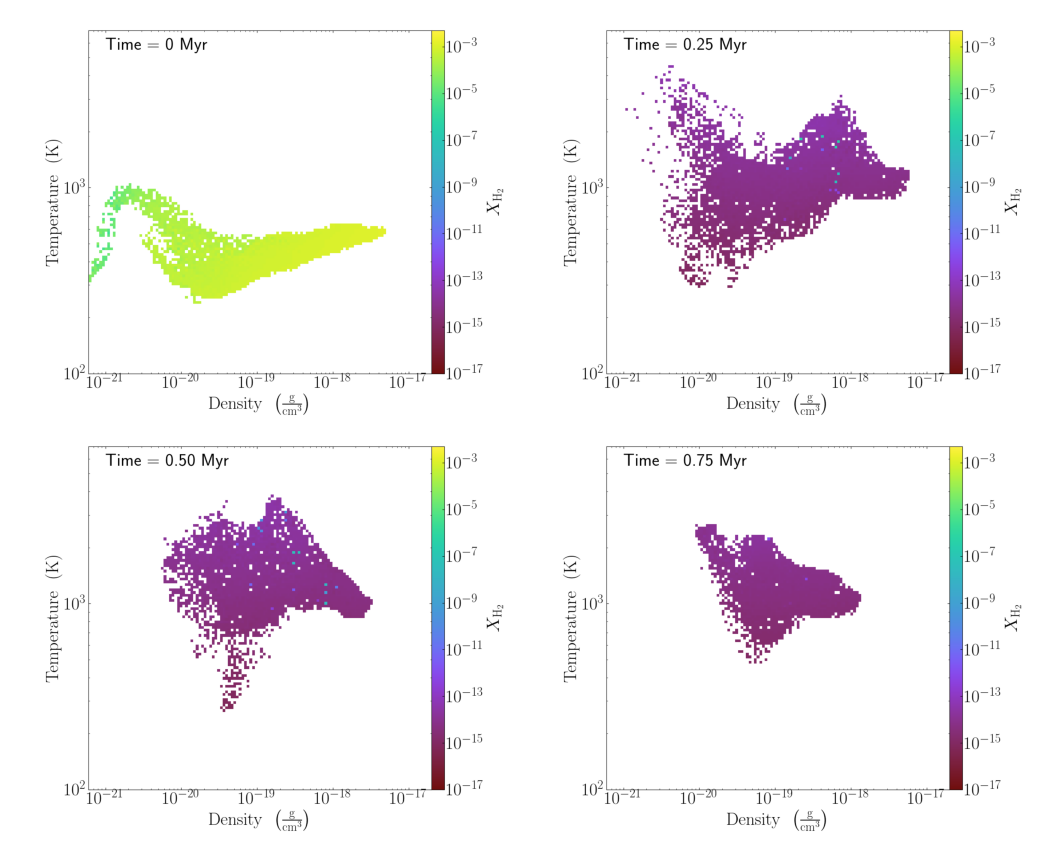}

    \caption{Phase diagrams of temperature, density, and $X_{\text{H}_2}$ surrounding sink A1 at $t = 0 \, \rm Myr$ (top left), $0.25 \, \rm Myr$ (top right), $0.50 \, \rm Myr$ (bottom left), and $0.75 \, \rm Myr$ (bottom right) in the no-cooling run. Each panel shows gas within 1 pc from sink A1. We see the gas temperature rises when the LW flux is added. This corresponds with $X_{\text{H}_2}$ dropping $\sim$ 10 orders of magnitude. The spread of temperatures decreases after the initial burst in heating, producing a more uniform high-temperature region. The gas clump also becomes more uniform in density. We see that the gas cloud heats up before settling into isothermal, quasistatic equilibrium at a temperature around $10^3 \, \rm K$, well below the atomic-cooling threshold.}
    \label{fig:1e10PhasePlot}
\end{figure*}

We show phase diagrams of the gas temperature, density, and $X_{\text{H}_2}$ to further reinforce these points (Fig.~\ref{fig:1e10PhasePlot}). At $t=0$, almost all the gas is below 1000 K. There is a wider range of densities ($5 \times 10^{-22} \,\rm g \, cm^{-3} \lesssim \rho \lesssim 5 \times10^{-18} \, g \, cm^{-3}$). The higher density gas has correspondingly higher $X_{\text{H}_2}$ ($\sim$$10^{-3}$) and the lower density gas has a lower value ($\sim$$10^{-5}$). $X_{\text{H}_2}$ quickly drops by roughly 10 orders of magnitude after the internal LW flux is turned on. The temperature experiences a corresponding increase while still retaining a wide spread in value ($t = 0.25 \, \rm Myr$, Fig.~\ref{fig:1e10PhasePlot}). This spread decreases with time, centering about the elevated average temperature. Apart from the initial spread in the range of temperatures, the phase space generally shrinks with time. The gas clump becomes increasingly uniform, centered around the average temperature and density displayed in Fig.~\ref{fig:1e10AvgTempDens}. The top row of Fig.~\ref{fig:1e10PhasePlot} displays the dynamical heating of the gas. The $\rm H_2$ is removed and the gas has begun to heat. However, the bottom row displays that this gas reaches thermal equilibrium below the atomic cooling threshold. We conclude that a massive star does not form even in this extreme scenario. 

We plot the halo virial temperature versus the radial distance from sink A1 to confirm the gas will not sufficiently heat to form a SMS. We define the virial temperature as $T_\text{vir} = \frac{1}{3}\frac{GM\mu m_\text{p}}{k_\text{B}r}$, where $M$ is the mass enclosed within radius $r$, $\mu$ is the mean molecular weight, and $m_\text{p}$ is the mass of proton. The potential energy per unit mass at radius $r$ is $-\frac{GM}{r}$, and the thermal energy per particle is $\frac{3}{2} k_\text{B}T$. This gives us the factor $1/3$. The LW radiation produces an increase in this virial temperature at low $r$, where the gas density dominates over dark matter in setting the virial temperature. However, it only manages to raise the virial temperature to $\sim$$1300 \,\rm K$. This agrees with the increase in temperature displayed in Figs.~\ref{fig:1e10PointTempDens} and~\ref{fig:1e10AvgTempDens}. Additionally, we find that $T_\text{gas} \sim T_\text{vir}$ (Fig.~\ref{fig:1e10AvgTempDens}), meaning that the gas has settled into quasi-hydrostatic equilibrium at this relatively low temperature. The gas will remain in this quasi-hydrostatic equilibrium until higher temperature gas at large radii begins falling into the halo center. However, the virial temperature does not approach the atomic cooling threshold until as far away as $r \gtrsim 100 \rm \, pc$. This is in part due to the halo's shallow DM profile. The gas's contribution to the virial temperature outweighs the DM's out to $\sim$$50 \, \rm pc$. The dynamical time for the gas near the atomic cooling threshold at $T \simeq 8000 \rm \, K$ is $\sim$$50 \rm \, Myr$. Any protostar will contract and join the main sequence due to the sustained periods of low accretion well before this high-temperature gas can collapse in the central regions. 

\begin{figure*}
    \centering
    \includegraphics[width=0.95\textwidth]{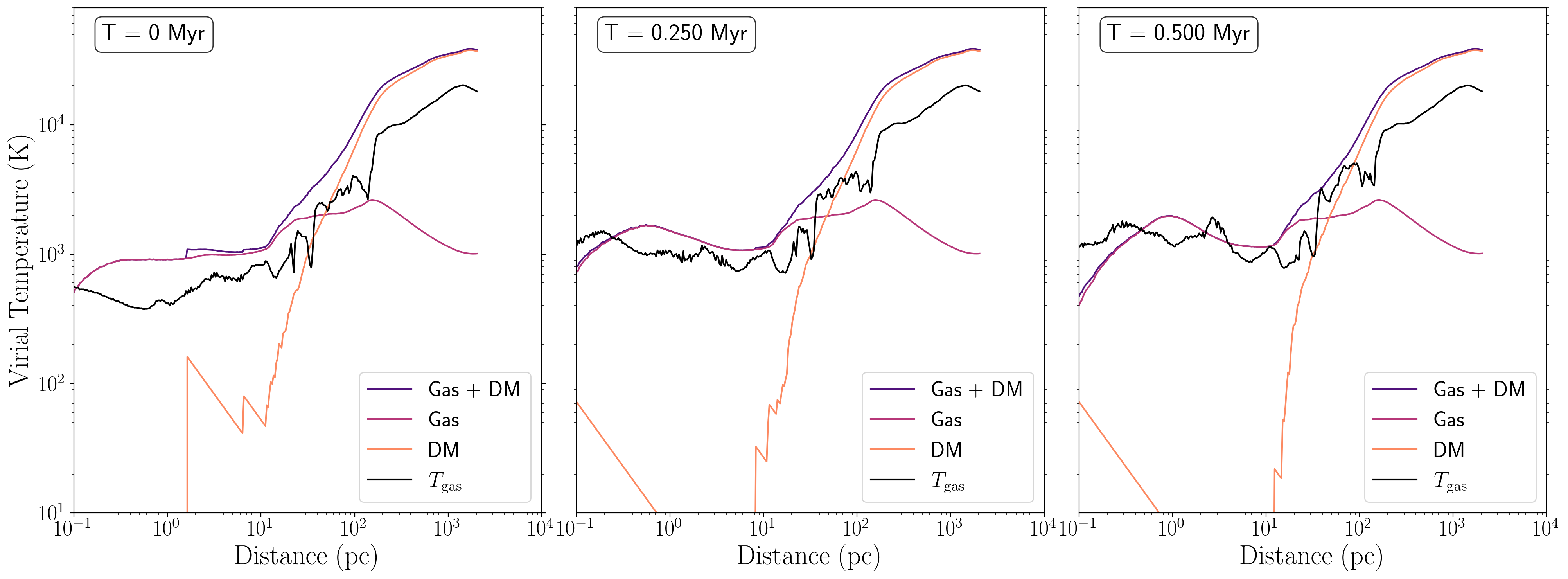}

    \caption{Radial profile of the virial temperature, centered around sink A1 in the no-cooling run. From left to right, the three panels show the profiles at snapshot \#10 (z = 6.5648), 0.25 Myr later at snapshot \#35 (z = 6.5633), and an additional 0.25 Myr later at snapshot \#60 (z = 6.5618). The gas and DM-only contributions to the virial temperature are shown alongside the actual virial temperature to display which component dominates at different radii. This demonstrates that the LW flux does not increase the virial temperature up to the atomic cooling threshold at small radii. The $T \simeq 10^4 \, \rm K$ gas at $\sim$$100 \, \rm pc$ has a dynamical time $t_\text{dyn} \simeq 50 \, \rm Myr$. It will not fall into the central clump in time to feed massive star formation.}
    \label{fig:TopVirialTemp}
\end{figure*}

\section{Conclusions}
\label{sec:Conclusions}

We ran a set of hydrodynamic simulations using \texttt{ENZO} on an ACH with two collapsing protostellar clouds. We added a large ``internal'' LW flux to model the protostellar feedback from the first protostar to form. We then measured the effects of this ``internal'' LW flux on the final protostellar masses and their local environment. We first determined that the expected LW flux from the first protostar is too weak to dissociate $\rm H_2$ around the second protostellar clump. This prevents the second protostar from forming through the collapse of warm, atomic hydrogen, instead producing a ``normal'' mass Pop III star through molecular cooling. Most significantly, we then demonstrated that the second protostellar clump will not produce a massive star even if the $\rm H_2$ is fully dissociated.  

The first result was determined through the background-only, short-delay, and long-delay runs (Section~\ref{sec:Results}). The additional LW flux produced a decrease in the accretion rates and final masses of the most massive protostars in each clump. In clump A, the final mass decreased from $7000 \, \rm M_\odot$ (background only) to $6000 \, \rm M_\odot$ for the long-delay and $4600 \, \rm M_\odot$ for the short-delay run. In clump B, the maximum mass measured decreased from $2000 \, \rm M_\odot$ (background only) to $1000 \rm \, M_\odot$ for the long-delay and $250 \, \rm M_\odot$ for the short-delay run. In each case, the accretion rates are below the critical accretion rate to avoid joining the main sequence. In clump A, this is due to the regions of increased temperature and decreased density around the sink particles. In clump B, we see the same trend, albeit with a higher central density at certain times. 

We found that even with $\rm H_2$ fully dissociated in the no-cooling run (Section~\ref{subsec:NoCooling}), the gas would not dynamically heat to the atomic cooling threshold within the simulated time. Instead, the gas became pressure supported, settling at a temperature of $\sim$$1200 \, \rm K$. This internal LW flux from sequential star formation can therefore not produce a massive star in this halo. Even if $\rm H_2$ is fully dissociated, there is not sufficient time for the gas to heat to the atomic cooling threshold. In the typical massive Pop III star formation route, this heating is driven by gas falling in from the virial radius. The large potential energy at the edges of the DM halo is converted into thermal energy as the gas falls in. The gas in our halo has already undergone this dynamical collapse and is no longer able to rely on it to heat itself. It instead settles into a quasi-static thermal equilibrium at $\sim$$1300 \, \rm K$, demonstrated by the similar gas and virial temperatures at small radii (Fig.~\ref{fig:1e10AvgTempDens}). It would require waiting $\sim$$50 \, \rm Myr$ for gas near the atomic-cooling threshold at $\sim$$100 \, \rm pc$ (Fig.~\ref{fig:TopVirialTemp}), to collapse. The Kelvin-Helmholtz contraction time, $\sim$$10^{4-5} \rm \, yr$, is orders of magnitude shorter. The protostar will therefore settle onto the main sequence long before the surrounding gas heats to the atomic cooling threshold. This makes it unlikely we will form a SMS within the second protostellar core.

We therefore find that our halo lacks two necessary conditions for forming a massive protostar. First, the LW flux is insufficient to dissociate the $\rm H_2$ within the protostellar cores. This could be resolved in halos with cores forming closer together, additional cores forming in quick succession, or a more massive first protostar. Each scenario would boost the ``internal'' LW flux, increasing the $\rm H_2$ survival density. For example, moving the halos to a distance $\sim$$5.78 \rm \, pc$ apart increases the $\rm H_2$ survival density to ${n_\text{H}} \cong 6.0 \times 10^6 \, \rm cm^{-3}$, surpassing the higher density regions of each clump. This would correspond to a new LW flux an order of magnitude larger ($\sim$$1.3 \times 10^5 \rm \,J_{21}$). The likelihood of scenarios such as forming clumps at much closer distances would need to be assessed. The second necessary piece is a deeper potential well. Dissociating the $\rm H_2$ in the ``no-cooling run'' shut off molecular cooling. The dynamical time in our halo, even with $\rm H_2$ is dissociated, is greater than $10^5 \, \rm yr$. This shut off accretion onto sink A1 and prevented additional sink particles, including sink B1, from forming. We need to decrease the dynamical time below this threshold in order to maintain protostar accretion rates $\gtrsim 0.01-0.04 \, \rm M_\odot \,yr^{-1}$. Again, periods of lower accretion lasting $\gtrsim 10^5 \, \rm yr$ produce a lower mass Pop III star. This necessary deeper potential well limits the scenario to more unique halos/regions. This makes the neighboring emission scenario unlikely to produce a large number of the massive BH seeds that later become SMBHs.

We expect these results to hold even more true for the more typical ACH population. As mentioned in \S~\ref{sec:SimulationSetup}, we analyzed an ACH that forms later and is more massive than ``typical'' ACHs, because its collapse was delayed by ionizing UV radiation. However, if anything, we expect these differences to make SMS formation easier. A large halo should have a deeper overall potential well, potentially increasing the gas accretion rate into the halo core. On the other hand, we do not expect a strong redshift dependence of the inner halo's potential. We therefore regard our result as conservative, and expect our conclusions to hold for higher-redshift ACHs. 

Ultimately, we conclude that neighboring protostellar cores do not help produce a massive star in either clump. However, they create a region of high-temperature gas around each of the protostellar cores. This could create a potential scenario in which a normal Pop III star forms within this well of relatively warm gas, which then undergoes rapid collapse at later times. The star may be unable to prevent the rapid collapse of this large $r$ gas cloud tens of Myrs later. This scenario could be possible within massive ACHs with a single star-forming core, driven by radiation from the first star itself, rather than a neighboring one, and will be the subject of a follow-up study.

\section{Acknowledgments}
\label{sec:Acknowledgements}
We acknowledge support from NSF grant AST-2006176 (ZH) as well as NSF grant AST-2009309, NASA ATP grant 80NSSC22K0629, and STScI grant JWST-AR-05238 (EV). The simulations in this work were run on Texas Advanced Computing Center's Stampede2 and Stampede3 systems. We used Stampede2 and Purdue University's computing system Anvil for data analysis. 







\bibliographystyle{mnras}
\bibliography{FirstYearReport10_2_24} 

\begin{thebibliography}{}
\makeatletter
\relax
\def\mn@urlcharsother{\let\do\@makeother \do\$\do\&\do\#\do\^\do\_\do\%\do\~}
\def\mn@doi{\begingroup\mn@urlcharsother \@ifnextchar [ {\mn@doi@} {\mn@doi@[]}}
\def\mn@doi@[#1]#2{\def\@tempa{#1}\ifx\@tempa\@empty \href {http://dx.doi.org/#2} {doi:#2}\else \href {http://dx.doi.org/#2} {#1}\fi \endgroup}
\def\mn@eprint#1#2{\mn@eprint@#1:#2::\@nil}
\def\mn@eprint@arXiv#1{\href {http://arxiv.org/abs/#1} {{\tt arXiv:#1}}}
\def\mn@eprint@dblp#1{\href {http://dblp.uni-trier.de/rec/bibtex/#1.xml} {dblp:#1}}
\def\mn@eprint@#1:#2:#3:#4\@nil{\def\@tempa {#1}\def\@tempb {#2}\def\@tempc {#3}\ifx \@tempc \@empty \let \@tempc \@tempb \let \@tempb \@tempa \fi \ifx \@tempb \@empty \def\@tempb {arXiv}\fi \@ifundefined {mn@eprint@\@tempb}{\@tempb:\@tempc}{\expandafter \expandafter \csname mn@eprint@\@tempb\endcsname \expandafter{\@tempc}}}

\bibitem[\protect\citeauthoryear{{Abel}, {Anninos}, {Zhang}  \& {Norman}}{{Abel} et~al.}{1997}]{Abel97}
{Abel} T.,  {Anninos} P.,  {Zhang} Y.,   {Norman} M.~L.,  1997, \mn@doi [\na] {10.1016/S1384-1076(97)00010-9}, \href {https://ui.adsabs.harvard.edu/abs/1997NewA....2..181A} {2, 181}

\bibitem[\protect\citeauthoryear{{Abel}, {Bryan}  \& {Norman}}{{Abel} et~al.}{2000}]{Abel00}
{Abel} T.,  {Bryan} G.~L.,   {Norman} M.~L.,  2000, \mn@doi [\apj] {10.1086/309295}, \href {https://ui.adsabs.harvard.edu/abs/2000ApJ...540...39A} {540, 39}

\bibitem[\protect\citeauthoryear{{Abel}, {Bryan}  \& {Norman}}{{Abel} et~al.}{2002}]{Abel02}
{Abel} T.,  {Bryan} G.~L.,   {Norman} M.~L.,  2002, \mn@doi [Science] {10.1126/science.295.5552.93}, \href {https://ui.adsabs.harvard.edu/abs/2002Sci...295...93A} {295, 93}

\bibitem[\protect\citeauthoryear{{Ahn}, {Shapiro}, {Iliev}, {Mellema}  \& {Pen}}{{Ahn} et~al.}{2009}]{Ahn09}
{Ahn} K.,  {Shapiro} P.~R.,  {Iliev} I.~T.,  {Mellema} G.,   {Pen} U.-L.,  2009, \mn@doi [\apj] {10.1088/0004-637X/695/2/1430}, \href {https://ui.adsabs.harvard.edu/abs/2009ApJ...695.1430A} {695, 1430}

\bibitem[\protect\citeauthoryear{{Alvarez}, {Wise}  \& {Abel}}{{Alvarez} et~al.}{2009}]{Alvarez09}
{Alvarez} M.~A.,  {Wise} J.~H.,   {Abel} T.,  2009, \mn@doi [\apjl] {10.1088/0004-637X/701/2/L133}, \href {https://ui.adsabs.harvard.edu/abs/2009ApJ...701L.133A} {701, L133}

\bibitem[\protect\citeauthoryear{{Ba{\~n}ados} et~al.,}{{Ba{\~n}ados} et~al.}{2018}]{Banados18}
{Ba{\~n}ados} E.,  et~al., 2018, \mn@doi [\nat] {10.1038/nature25180}, \href {https://ui.adsabs.harvard.edu/abs/2018Natur.553..473B} {553, 473}

\bibitem[\protect\citeauthoryear{Becerra, Greif, Springel  \& Hernquist}{Becerra et~al.}{2014}]{Becerra14}
Becerra F.,  Greif T.~H.,  Springel V.,   Hernquist L.~E.,  2014, \mn@doi [Monthly Notices of the Royal Astronomical Society] {10.1093/mnras/stu2284}, 446, 2380

\bibitem[\protect\citeauthoryear{{Begelman} \& {Rees}}{{Begelman} \& {Rees}}{1978}]{Begelman78}
{Begelman} M.~C.,  {Rees} M.~J.,  1978, \mn@doi [\mnras] {10.1093/mnras/185.4.847}, \href {https://ui.adsabs.harvard.edu/abs/1978MNRAS.185..847B} {185, 847}

\bibitem[\protect\citeauthoryear{{Begelman}, {Volonteri}  \& {Rees}}{{Begelman} et~al.}{2006}]{Begelman06}
{Begelman} M.~C.,  {Volonteri} M.,   {Rees} M.~J.,  2006, \mn@doi [\mnras] {10.1111/j.1365-2966.2006.10467.x}, \href {https://ui.adsabs.harvard.edu/abs/2006MNRAS.370..289B} {370, 289}

\bibitem[\protect\citeauthoryear{{Begelman}, {Rossi}  \& {Armitage}}{{Begelman} et~al.}{2008}]{Begelman08}
{Begelman} M.~C.,  {Rossi} E.~M.,   {Armitage} P.~J.,  2008, \mn@doi [\mnras] {10.1111/j.1365-2966.2008.13344.x}, \href {https://ui.adsabs.harvard.edu/abs/2008MNRAS.387.1649B} {387, 1649}

\bibitem[\protect\citeauthoryear{{Berger} \& {Colella}}{{Berger} \& {Colella}}{1989}]{Berger89}
{Berger} M.~J.,  {Colella} P.,  1989, \mn@doi [Journal of Computational Physics] {10.1016/0021-9991(89)90035-1}, \href {https://ui.adsabs.harvard.edu/abs/1989JCoPh..82...64B} {82, 64}

\bibitem[\protect\citeauthoryear{{Bosman}}{{Bosman}}{2024}]{Bosman24}
{Bosman} S.,  2024, {All z>5.7 quasars currently known}, Zenodo dataset, \mn@doi{https://doi.org/10.5281/zenodo.13170305}

\bibitem[\protect\citeauthoryear{{Bromm}}{{Bromm}}{2013}]{Bromm13}
{Bromm} V.,  2013, \mn@doi [Reports on Progress in Physics] {10.1088/0034-4885/76/11/112901}, \href {https://ui.adsabs.harvard.edu/abs/2013RPPh...76k2901B} {76, 112901}

\bibitem[\protect\citeauthoryear{Bromm \& Loeb}{Bromm \& Loeb}{2003}]{Bromm03}
Bromm V.,  Loeb A.,  2003, \mn@doi [The Astrophysical Journal] {10.1086/377529}, 596, 34

\bibitem[\protect\citeauthoryear{Bromm \& Yoshida}{Bromm \& Yoshida}{2011}]{Bromm11}
Bromm V.,  Yoshida N.,  2011, \mn@doi [Annual Review of Astronomy and Astrophysics] {10.1146/annurev-astro-081710-102608}, 49, 373

\bibitem[\protect\citeauthoryear{Bromm, Coppi  \& Larson}{Bromm et~al.}{2002}]{Bromm02}
Bromm V.,  Coppi P.~S.,   Larson R.~B.,  2002, \mn@doi [The Astrophysical Journal] {10.1086/323947}, 564, 23

\bibitem[\protect\citeauthoryear{{Bryan} \& {Norman}}{{Bryan} \& {Norman}}{1997}]{Bryan97}
{Bryan} G.~L.,  {Norman} M.~L.,  1997, \mn@doi [arXiv e-prints] {10.48550/arXiv.astro-ph/9710187}, \href {https://ui.adsabs.harvard.edu/abs/1997astro.ph.10187B} {pp astro--ph/9710187}

\bibitem[\protect\citeauthoryear{{Bryan} et~al.,}{{Bryan} et~al.}{2014}]{Bryan14}
{Bryan} G.~L.,  et~al., 2014, \mn@doi [\apjs] {10.1088/0067-0049/211/2/19}, \href {https://ui.adsabs.harvard.edu/abs/2014ApJS..211...19B} {211, 19}

\bibitem[\protect\citeauthoryear{{Chon} \& {Latif}}{{Chon} \& {Latif}}{2017}]{Chon17}
{Chon} S.,  {Latif} M.~A.,  2017, \mn@doi [\mnras] {10.1093/mnras/stx348}, \href {https://ui.adsabs.harvard.edu/abs/2017MNRAS.467.4293C} {467, 4293}

\bibitem[\protect\citeauthoryear{{Devecchi} \& {Volonteri}}{{Devecchi} \& {Volonteri}}{2009}]{Devecch09}
{Devecchi} B.,  {Volonteri} M.,  2009, \mn@doi [\apj] {10.1088/0004-637X/694/1/302}, \href {https://ui.adsabs.harvard.edu/abs/2009ApJ...694..302D} {694, 302}

\bibitem[\protect\citeauthoryear{{Dijkstra}, {Haiman}, {Mesinger}  \& {Wyithe}}{{Dijkstra} et~al.}{2008}]{Dijkstra08}
{Dijkstra} M.,  {Haiman} Z.,  {Mesinger} A.,   {Wyithe} J. S.~B.,  2008, \mn@doi [\mnras] {10.1111/j.1365-2966.2008.14031.x}, \href {https://ui.adsabs.harvard.edu/abs/2008MNRAS.391.1961D} {391, 1961}

\bibitem[\protect\citeauthoryear{{Dunn}, {Bellovary}, {Holley-Bockelmann}, {Christensen}  \& {Quinn}}{{Dunn} et~al.}{2018}]{Dunn18}
{Dunn} G.,  {Bellovary} J.,  {Holley-Bockelmann} K.,  {Christensen} C.,   {Quinn} T.,  2018, \mn@doi [\apj] {10.3847/1538-4357/aac7c2}, \href {https://ui.adsabs.harvard.edu/abs/2018ApJ...861...39D} {861, 39}

\bibitem[\protect\citeauthoryear{{Fan} et~al.,}{{Fan} et~al.}{2006}]{Fan06}
{Fan} X.,  et~al., 2006, \mn@doi [\aj] {10.1086/500296}, \href {https://ui.adsabs.harvard.edu/abs/2006AJ....131.1203F} {131, 1203}

\bibitem[\protect\citeauthoryear{{Fan}, {Ba{\~n}ados}  \& {Simcoe}}{{Fan} et~al.}{2023}]{Fan23}
{Fan} X.,  {Ba{\~n}ados} E.,   {Simcoe} R.~A.,  2023, \mn@doi [\araa] {10.1146/annurev-astro-052920-102455}, \href {https://ui.adsabs.harvard.edu/abs/2023ARA&A..61..373F} {61, 373}

\bibitem[\protect\citeauthoryear{{Feathers}, {Visbal}, {Kulkarni}  \& {Hazlett}}{{Feathers} et~al.}{2023}]{Feathers23}
{Feathers} C.~R.,  {Visbal} E.,  {Kulkarni} M.,   {Hazlett} R.,  2023, \mn@doi [arXiv e-prints] {10.48550/arXiv.2306.07371}, \href {https://ui.adsabs.harvard.edu/abs/2023arXiv230607371F} {p. arXiv:2306.07371}

\bibitem[\protect\citeauthoryear{{Galli} \& {Palla}}{{Galli} \& {Palla}}{1998}]{Galli98}
{Galli} D.,  {Palla} F.,  1998, \mn@doi [\aap] {10.48550/arXiv.astro-ph/9803315}, \href {https://ui.adsabs.harvard.edu/abs/1998A&A...335..403G} {335, 403}

\bibitem[\protect\citeauthoryear{{Gonz{\'a}lez}, {Kremer}, {Chatterjee}, {Fragione}, {Rodriguez}, {Weatherford}, {Ye}  \& {Rasio}}{{Gonz{\'a}lez} et~al.}{2021}]{Gonzalez21}
{Gonz{\'a}lez} E.,  {Kremer} K.,  {Chatterjee} S.,  {Fragione} G.,  {Rodriguez} C.~L.,  {Weatherford} N.~C.,  {Ye} C.~S.,   {Rasio} F.~A.,  2021, \mn@doi [\apjl] {10.3847/2041-8213/abdf5b}, \href {https://ui.adsabs.harvard.edu/abs/2021ApJ...908L..29G} {908, L29}

\bibitem[\protect\citeauthoryear{{Greif}}{{Greif}}{2015}]{Greif15}
{Greif} T.~H.,  2015, \mn@doi [Computational Astrophysics and Cosmology] {10.1186/s40668-014-0006-2}, \href {https://ui.adsabs.harvard.edu/abs/2015ComAC...2....3G} {2, 3}

\bibitem[\protect\citeauthoryear{{Greif}, {Bromm}, {Clark}, {Glover}, {Smith}, {Klessen}, {Yoshida}  \& {Springel}}{{Greif} et~al.}{2012}]{Grief12}
{Greif} T.~H.,  {Bromm} V.,  {Clark} P.~C.,  {Glover} S. C.~O.,  {Smith} R.~J.,  {Klessen} R.~S.,  {Yoshida} N.,   {Springel} V.,  2012, in {Umemura} M.,  {Omukai} K.,  eds,  American Institute of Physics Conference Series Vol. 1480, First Stars IV - from Hayashi to the Future -. pp 51--56 (\mn@eprint {arXiv} {1202.5552}), \mn@doi{10.1063/1.4754327}

\bibitem[\protect\citeauthoryear{{Haemmerl{\'e}}, {Woods}, {Klessen}, {Heger}  \& {Whalen}}{{Haemmerl{\'e}} et~al.}{2018}]{Haemmerle18}
{Haemmerl{\'e}} L.,  {Woods} T.~E.,  {Klessen} R.~S.,  {Heger} A.,   {Whalen} D.~J.,  2018, \mn@doi [\mnras] {10.1093/mnras/stx2919}, \href {https://ui.adsabs.harvard.edu/abs/2018MNRAS.474.2757H} {474, 2757}

\bibitem[\protect\citeauthoryear{Haiman \& Loeb}{Haiman \& Loeb}{1997}]{Haiman97}
Haiman Z.,  Loeb A.,  1997, \mn@doi [The Astrophysical Journal] {10.1086/304238}, 483, 21

\bibitem[\protect\citeauthoryear{{Haiman}, {Thoul}  \& {Loeb}}{{Haiman} et~al.}{1996}]{Haiman96}
{Haiman} Z.,  {Thoul} A.~A.,   {Loeb} A.,  1996, \mn@doi [\apj] {10.1086/177343}, \href {https://ui.adsabs.harvard.edu/abs/1996ApJ...464..523H} {464, 523}

\bibitem[\protect\citeauthoryear{{Haiman}, {Abel}  \& {Rees}}{{Haiman} et~al.}{2000}]{Haiman00}
{Haiman} Z.,  {Abel} T.,   {Rees} M.~J.,  2000, \mn@doi [\apj] {10.1086/308723}, \href {https://ui.adsabs.harvard.edu/abs/2000ApJ...534...11H} {534, 11}

\bibitem[\protect\citeauthoryear{{Hirano}, {Hosokawa}, {Yoshida}, {Umeda}, {Omukai}, {Chiaki}  \& {Yorke}}{{Hirano} et~al.}{2014}]{Hirano14}
{Hirano} S.,  {Hosokawa} T.,  {Yoshida} N.,  {Umeda} H.,  {Omukai} K.,  {Chiaki} G.,   {Yorke} H.~W.,  2014, \mn@doi [\apj] {10.1088/0004-637X/781/2/60}, \href {https://ui.adsabs.harvard.edu/abs/2014ApJ...781...60H} {781, 60}

\bibitem[\protect\citeauthoryear{{Hosokawa}, {Omukai}  \& {Yorke}}{{Hosokawa} et~al.}{2012}]{Hosokawa12}
{Hosokawa} T.,  {Omukai} K.,   {Yorke} H.~W.,  2012, \mn@doi [\apj] {10.1088/0004-637X/756/1/93}, \href {https://ui.adsabs.harvard.edu/abs/2012ApJ...756...93H} {756, 93}

\bibitem[\protect\citeauthoryear{{Hosokawa}, {Yorke}, {Inayoshi}, {Omukai}  \& {Yoshida}}{{Hosokawa} et~al.}{2013}]{Hosokawa13b}
{Hosokawa} T.,  {Yorke} H.~W.,  {Inayoshi} K.,  {Omukai} K.,   {Yoshida} N.,  2013, \mn@doi [\apj] {10.1088/0004-637X/778/2/178}, \href {https://ui.adsabs.harvard.edu/abs/2013ApJ...778..178H} {778, 178}

\bibitem[\protect\citeauthoryear{{Inayoshi}, {Visbal}  \& {Haiman}}{{Inayoshi} et~al.}{2020}]{Inayoshi20}
{Inayoshi} K.,  {Visbal} E.,   {Haiman} Z.,  2020, \mn@doi [\araa] {10.1146/annurev-astro-120419-014455}, \href {https://ui.adsabs.harvard.edu/abs/2020ARA&A..58...27I} {58, 27}

\bibitem[\protect\citeauthoryear{{Katz}, {Sijacki}  \& {Haehnelt}}{{Katz} et~al.}{2015}]{Katz15}
{Katz} H.,  {Sijacki} D.,   {Haehnelt} M.~G.,  2015, \mn@doi [\mnras] {10.1093/mnras/stv1048}, \href {https://ui.adsabs.harvard.edu/abs/2015MNRAS.451.2352K} {451, 2352}

\bibitem[\protect\citeauthoryear{{Klessen}}{{Klessen}}{2019}]{Klessen19}
{Klessen} R.,  2019, in {Latif} M.,  {Schleicher} D.,  eds, , Formation of the First Black Holes.
pp 67--97, \mn@doi{10.1142/9789813227958_0004}

\bibitem[\protect\citeauthoryear{{Klessen} \& {Glover}}{{Klessen} \& {Glover}}{2023}]{Klessen23}
{Klessen} R.~S.,  {Glover} S. C.~O.,  2023, \mn@doi [\araa] {10.1146/annurev-astro-071221-053453}, \href {https://ui.adsabs.harvard.edu/abs/2023ARA&A..61...65K} {61, 65}

\bibitem[\protect\citeauthoryear{{Kormendy} \& {Ho}}{{Kormendy} \& {Ho}}{2013}]{Kormendy13}
{Kormendy} J.,  {Ho} L.~C.,  2013, \mn@doi [\araa] {10.1146/annurev-astro-082708-101811}, \href {https://ui.adsabs.harvard.edu/abs/2013ARA&A..51..511K} {51, 511}

\bibitem[\protect\citeauthoryear{{Krumholz}, {McKee}  \& {Klein}}{{Krumholz} et~al.}{2004}]{Krumholz04}
{Krumholz} M.~R.,  {McKee} C.~F.,   {Klein} R.~I.,  2004, \mn@doi [\apj] {10.1086/421935}, \href {https://ui.adsabs.harvard.edu/abs/2004ApJ...611..399K} {611, 399}

\bibitem[\protect\citeauthoryear{{Kulkarni}, {Visbal}  \& {Bryan}}{{Kulkarni} et~al.}{2019}]{Kulkarni19}
{Kulkarni} M.,  {Visbal} E.,   {Bryan} G.~L.,  2019, \mn@doi [\apj] {10.3847/1538-4357/ab35e2}, \href {https://ui.adsabs.harvard.edu/abs/2019ApJ...882..178K} {882, 178}

\bibitem[\protect\citeauthoryear{{Kulkarni}, {Visbal}  \& {Bryan}}{{Kulkarni} et~al.}{2021}]{Kulkarni21}
{Kulkarni} M.,  {Visbal} E.,   {Bryan} G.~L.,  2021, \mn@doi [\apj] {10.3847/1538-4357/ac08a3}, \href {https://ui.adsabs.harvard.edu/abs/2021ApJ...917...40K} {917, 40}

\bibitem[\protect\citeauthoryear{{Latif}, {Schleicher}  \& {Hartwig}}{{Latif} et~al.}{2016}]{Latif16}
{Latif} M.~A.,  {Schleicher} D.~R.~G.,   {Hartwig} T.,  2016, \mn@doi [\mnras] {10.1093/mnras/stw297}, \href {https://ui.adsabs.harvard.edu/abs/2016MNRAS.458..233L} {458, 233}

\bibitem[\protect\citeauthoryear{Machacek, Bryan  \& Abel}{Machacek et~al.}{2001}]{Machacek01}
Machacek M.~E.,  Bryan G.~L.,   Abel T.,  2001, \mn@doi [The Astrophysical Journal] {10.1086/319014}, 548, 509

\bibitem[\protect\citeauthoryear{{McGreer} \& {Bryan}}{{McGreer} \& {Bryan}}{2008}]{McGreer08}
{McGreer} I.~D.,  {Bryan} G.~L.,  2008, \mn@doi [\apj] {10.1086/590530}, \href {https://ui.adsabs.harvard.edu/abs/2008ApJ...685....8M} {685, 8}

\bibitem[\protect\citeauthoryear{{Milosavljevi{\'c}}, {Couch}  \& {Bromm}}{{Milosavljevi{\'c}} et~al.}{2009}]{Milo09}
{Milosavljevi{\'c}} M.,  {Couch} S.~M.,   {Bromm} V.,  2009, \mn@doi [\apjl] {10.1088/0004-637X/696/2/L146}, \href {https://ui.adsabs.harvard.edu/abs/2009ApJ...696L.146M} {696, L146}

\bibitem[\protect\citeauthoryear{{Nandal}, {Regan}, {Woods}, {Farrell}, {Ekstr{\"o}m}  \& {Meynet}}{{Nandal} et~al.}{2023}]{Nandal23}
{Nandal} D.,  {Regan} J.~A.,  {Woods} T.~E.,  {Farrell} E.,  {Ekstr{\"o}m} S.,   {Meynet} G.,  2023, \mn@doi [arXiv e-prints] {10.48550/arXiv.2306.17223}, \href {https://ui.adsabs.harvard.edu/abs/2023arXiv230617223N} {p. arXiv:2306.17223}

\bibitem[\protect\citeauthoryear{{O'Shea} \& {Norman}}{{O'Shea} \& {Norman}}{2007}]{OShea07}
{O'Shea} B.~W.,  {Norman} M.~L.,  2007, \mn@doi [\apj] {10.1086/509250}, \href {https://ui.adsabs.harvard.edu/abs/2007ApJ...654...66O} {654, 66}

\bibitem[\protect\citeauthoryear{{O'Shea} \& {Norman}}{{O'Shea} \& {Norman}}{2008}]{OShea08}
{O'Shea} B.~W.,  {Norman} M.~L.,  2008, \mn@doi [\apj] {10.1086/524006}, \href {https://ui.adsabs.harvard.edu/abs/2008ApJ...673...14O} {673, 14}

\bibitem[\protect\citeauthoryear{{O'Shea}, {Bryan}, {Bordner}, {Norman}, {Abel}, {Harkness}  \& {Kritsuk}}{{O'Shea} et~al.}{2004}]{OShea04}
{O'Shea} B.~W.,  {Bryan} G.,  {Bordner} J.,  {Norman} M.~L.,  {Abel} T.,  {Harkness} R.,   {Kritsuk} A.,  2004, \mn@doi [arXiv e-prints] {10.48550/arXiv.astro-ph/0403044}, \href {https://ui.adsabs.harvard.edu/abs/2004astro.ph..3044O} {pp astro--ph/0403044}

\bibitem[\protect\citeauthoryear{{Omukai}}{{Omukai}}{2001}]{Omukai01}
{Omukai} K.,  2001, \mn@doi [\apj] {10.1086/318296}, \href {https://ui.adsabs.harvard.edu/abs/2001ApJ...546..635O} {546, 635}

\bibitem[\protect\citeauthoryear{{Omukai}, {Schneider}  \& {Haiman}}{{Omukai} et~al.}{2008}]{Omukai08}
{Omukai} K.,  {Schneider} R.,   {Haiman} Z.,  2008, \mn@doi [\apj] {10.1086/591636}, \href {https://ui.adsabs.harvard.edu/abs/2008ApJ...686..801O} {686, 801}

\bibitem[\protect\citeauthoryear{{Planck Collaboration} et~al.,}{{Planck Collaboration} et~al.}{2014}]{Planck14}
{Planck Collaboration} et~al., 2014, \mn@doi [A&A] {10.1051/0004-6361/201321591}, 571, A16

\bibitem[\protect\citeauthoryear{{Portegies Zwart}, {Baumgardt}, {Hut}, {Makino}  \& {McMillan}}{{Portegies Zwart} et~al.}{2004}]{Zwart04}
{Portegies Zwart} S.~F.,  {Baumgardt} H.,  {Hut} P.,  {Makino} J.,   {McMillan} S. L.~W.,  2004, \mn@doi [\nat] {10.1038/nature02448}, \href {https://ui.adsabs.harvard.edu/abs/2004Natur.428..724P} {428, 724}

\bibitem[\protect\citeauthoryear{Prieto, Jimenez  \& Haiman}{Prieto et~al.}{2013}]{Prieto13}
Prieto J.,  Jimenez R.,   Haiman Z.,  2013, \mn@doi [Monthly Notices of the Royal Astronomical Society] {10.1093/mnras/stt1730}, 436, 2301

\bibitem[\protect\citeauthoryear{{Rahmati}, {Pawlik}, {Rai{\v{c}}evi{\'c}}  \& {Schaye}}{{Rahmati} et~al.}{2013}]{Rahmati13}
{Rahmati} A.,  {Pawlik} A.~H.,  {Rai{\v{c}}evi{\'c}} M.,   {Schaye} J.,  2013, \mn@doi [\mnras] {10.1093/mnras/stt066}, \href {https://ui.adsabs.harvard.edu/abs/2013MNRAS.430.2427R} {430, 2427}

\bibitem[\protect\citeauthoryear{{Rees}}{{Rees}}{1978}]{Rees78}
{Rees} M.~J.,  1978, \mn@doi [\physscr] {10.1088/0031-8949/17/3/010}, \href {https://ui.adsabs.harvard.edu/abs/1978PhyS...17..193R} {17, 193}

\bibitem[\protect\citeauthoryear{{Regan} \& {Downes}}{{Regan} \& {Downes}}{2018}]{Regan18}
{Regan} J.~A.,  {Downes} T.~P.,  2018, \mn@doi [\mnras] {10.1093/mnras/sty1289}, \href {https://ui.adsabs.harvard.edu/abs/2018MNRAS.478.5037R} {478, 5037}

\bibitem[\protect\citeauthoryear{{Regan} \& {Haehnelt}}{{Regan} \& {Haehnelt}}{2009a}]{Regan09b}
{Regan} J.~A.,  {Haehnelt} M.~G.,  2009a, \mn@doi [\mnras] {10.1111/j.1365-2966.2008.14088.x}, \href {https://ui.adsabs.harvard.edu/abs/2009MNRAS.393..858R} {393, 858}

\bibitem[\protect\citeauthoryear{{Regan} \& {Haehnelt}}{{Regan} \& {Haehnelt}}{2009b}]{Regan09a}
{Regan} J.~A.,  {Haehnelt} M.~G.,  2009b, \mn@doi [\mnras] {10.1111/j.1365-2966.2009.14579.x}, \href {https://ui.adsabs.harvard.edu/abs/2009MNRAS.396..343R} {396, 343}

\bibitem[\protect\citeauthoryear{Regan, Visbal, Wise, Haiman, Johansson  \& Bryan}{Regan et~al.}{2017}]{Regan17}
Regan J.~A.,  Visbal E.,  Wise J.~H.,  Haiman Z.,  Johansson P.~H.,   Bryan G.~L.,  2017, \mn@doi [Nature Astronomy] {10.1038/s41550-017-0075}, 1, 0075

\bibitem[\protect\citeauthoryear{{Regan}, {Wise}, {O'Shea}  \& {Norman}}{{Regan} et~al.}{2020}]{Regan20}
{Regan} J.~A.,  {Wise} J.~H.,  {O'Shea} B.~W.,   {Norman} M.~L.,  2020, \mn@doi [\mnras] {10.1093/mnras/staa035}, \href {https://ui.adsabs.harvard.edu/abs/2020MNRAS.492.3021R} {492, 3021}

\bibitem[\protect\citeauthoryear{{Rizzuto} et~al.,}{{Rizzuto} et~al.}{2021}]{Rizzuto21}
{Rizzuto} F.~P.,  et~al., 2021, \mn@doi [\mnras] {10.1093/mnras/staa3634}, \href {https://ui.adsabs.harvard.edu/abs/2021MNRAS.501.5257R} {501, 5257}

\bibitem[\protect\citeauthoryear{{Sakurai}, {Vorobyov}, {Hosokawa}, {Yoshida}, {Omukai}  \& {Yorke}}{{Sakurai} et~al.}{2016}]{Sakurai16}
{Sakurai} Y.,  {Vorobyov} E.~I.,  {Hosokawa} T.,  {Yoshida} N.,  {Omukai} K.,   {Yorke} H.~W.,  2016, \mn@doi [\mnras] {10.1093/mnras/stw637}, \href {https://ui.adsabs.harvard.edu/abs/2016MNRAS.459.1137S} {459, 1137}

\bibitem[\protect\citeauthoryear{{Shang}, {Bryan}  \& {Haiman}}{{Shang} et~al.}{2010}]{Shang10}
{Shang} C.,  {Bryan} G.~L.,   {Haiman} Z.,  2010, \mn@doi [\mnras] {10.1111/j.1365-2966.2009.15960.x}, \href {https://ui.adsabs.harvard.edu/abs/2010MNRAS.402.1249S} {402, 1249}

\bibitem[\protect\citeauthoryear{{Smith}, {Regan}, {Downes}, {Norman}, {O'Shea}  \& {Wise}}{{Smith} et~al.}{2018}]{Smith18}
{Smith} B.~D.,  {Regan} J.~A.,  {Downes} T.~P.,  {Norman} M.~L.,  {O'Shea} B.~W.,   {Wise} J.~H.,  2018, \mn@doi [\mnras] {10.1093/mnras/sty2103}, \href {https://ui.adsabs.harvard.edu/abs/2018MNRAS.480.3762S} {480, 3762}

\bibitem[\protect\citeauthoryear{Spinoso, Bonoli, Valiante, Schneider  \& Izquierdo-Villalba}{Spinoso et~al.}{2022}]{Spinoso22}
Spinoso D.,  Bonoli S.,  Valiante R.,  Schneider R.,   Izquierdo-Villalba D.,  2022, \mn@doi [Monthly Notices of the Royal Astronomical Society] {10.1093/mnras/stac3169}, 518, 4672

\bibitem[\protect\citeauthoryear{{Stacy}}{{Stacy}}{2012}]{Stacy12}
{Stacy} A.,  2012, in American Astronomical Society Meeting Abstracts \#219. p. 444.05

\bibitem[\protect\citeauthoryear{{Stacy}, {Greif}  \& {Bromm}}{{Stacy} et~al.}{2010}]{Stacy10}
{Stacy} A.,  {Greif} T.~H.,   {Bromm} V.,  2010, \mn@doi [\mnras] {10.1111/j.1365-2966.2009.16113.x}, \href {https://ui.adsabs.harvard.edu/abs/2010MNRAS.403...45S} {403, 45}

\bibitem[\protect\citeauthoryear{{Sugimura}, {Omukai}  \& {Inoue}}{{Sugimura} et~al.}{2014}]{Sugimura14}
{Sugimura} K.,  {Omukai} K.,   {Inoue} A.~K.,  2014, \mn@doi [\mnras] {10.1093/mnras/stu1778}, \href {https://ui.adsabs.harvard.edu/abs/2014MNRAS.445..544S} {445, 544}

\bibitem[\protect\citeauthoryear{Tegmark, Silk, Rees, Blanchard, Abel  \& Palla}{Tegmark et~al.}{1997}]{Tegmark97}
Tegmark M.,  Silk J.,  Rees M.~J.,  Blanchard A.,  Abel T.,   Palla F.,  1997, \mn@doi [The Astrophysical Journal] {10.1086/303434}, 474, 1

\bibitem[\protect\citeauthoryear{{Venemans} et~al.,}{{Venemans} et~al.}{2015}]{Venemans15}
{Venemans} B.~P.,  et~al., 2015, \mn@doi [\apjl] {10.1088/2041-8205/801/1/L11}, \href {https://ui.adsabs.harvard.edu/abs/2015ApJ...801L..11V} {801, L11}

\bibitem[\protect\citeauthoryear{{Visbal}, {Haiman}  \& {Bryan}}{{Visbal} et~al.}{2014}]{Visbal14}
{Visbal} E.,  {Haiman} Z.,   {Bryan} G.~L.,  2014, \mn@doi [\mnras] {10.1093/mnras/stu1794}, \href {https://ui.adsabs.harvard.edu/abs/2014MNRAS.445.1056V} {445, 1056}

\bibitem[\protect\citeauthoryear{{Visbal}, {Bryan}  \& {Haiman}}{{Visbal} et~al.}{2017}]{Visbal17}
{Visbal} E.,  {Bryan} G.~L.,   {Haiman} Z.,  2017, \mn@doi [\mnras] {10.1093/mnras/stx909}, \href {https://ui.adsabs.harvard.edu/abs/2017MNRAS.469.1456V} {469, 1456}

\bibitem[\protect\citeauthoryear{{Volonteri}, {Habouzit}  \& {Colpi}}{{Volonteri} et~al.}{2021}]{Volonteri21}
{Volonteri} M.,  {Habouzit} M.,   {Colpi} M.,  2021, \mn@doi [Nature Reviews Physics] {10.1038/s42254-021-00364-9}, \href {https://ui.adsabs.harvard.edu/abs/2021NatRP...3..732V} {3, 732}

\bibitem[\protect\citeauthoryear{{Wang} et~al.,}{{Wang} et~al.}{2021}]{Wang21}
{Wang} F.,  et~al., 2021, \mn@doi [\apjl] {10.3847/2041-8213/abd8c6}, \href {https://ui.adsabs.harvard.edu/abs/2021ApJ...907L...1W} {907, L1}

\bibitem[\protect\citeauthoryear{Wise \& Abel}{Wise \& Abel}{2007}]{Wise07}
Wise J.~H.,  Abel T.,  2007, \mn@doi [The Astrophysical Journal] {10.1086/522876}, 671, 1559

\bibitem[\protect\citeauthoryear{Wise, Turk  \& Abel}{Wise et~al.}{2008}]{Wise08}
Wise J.~H.,  Turk M.~J.,   Abel T.,  2008, \mn@doi [The Astrophysical Journal] {10.1086/588209}, 682, 745

\bibitem[\protect\citeauthoryear{{Wolcott-Green} \& {Haiman}}{{Wolcott-Green} \& {Haiman}}{2019}]{Wolcott19}
{Wolcott-Green} J.,  {Haiman} Z.,  2019, \mn@doi [\mnras] {10.1093/mnras/sty3280}, \href {https://ui.adsabs.harvard.edu/abs/2019MNRAS.484.2467W} {484, 2467}

\bibitem[\protect\citeauthoryear{{Wolcott-Green}, {Haiman}  \& {Bryan}}{{Wolcott-Green} et~al.}{2017}]{Wolcott17}
{Wolcott-Green} J.,  {Haiman} Z.,   {Bryan} G.~L.,  2017, \mn@doi [\mnras] {10.1093/mnras/stx167}, \href {https://ui.adsabs.harvard.edu/abs/2017MNRAS.469.3329W} {469, 3329}

\bibitem[\protect\citeauthoryear{{Woods}, {Heger}, {Whalen}, {Haemmerl{\'e}}  \& {Klessen}}{{Woods} et~al.}{2017}]{Woods17}
{Woods} T.~E.,  {Heger} A.,  {Whalen} D.~J.,  {Haemmerl{\'e}} L.,   {Klessen} R.~S.,  2017, \mn@doi [\apjl] {10.3847/2041-8213/aa7412}, \href {https://ui.adsabs.harvard.edu/abs/2017ApJ...842L...6W} {842, L6}

\bibitem[\protect\citeauthoryear{{Woods} et~al.,}{{Woods} et~al.}{2019}]{Woods19}
{Woods} T.~E.,  et~al., 2019, \mn@doi [\pasa] {10.1017/pasa.2019.14}, \href {https://ui.adsabs.harvard.edu/abs/2019PASA...36...27W} {36, e027}

\bibitem[\protect\citeauthoryear{Wu et~al.,}{Wu et~al.}{2015}]{Wu_2015}
Wu X.-B.,  et~al., 2015, \mn@doi [Nature] {10.1038/nature14241}, 518, 512

\bibitem[\protect\citeauthoryear{Yoshida, Abel, Hernquist  \& Sugiyama}{Yoshida et~al.}{2003}]{Yoshida03}
Yoshida N.,  Abel T.,  Hernquist L.,   Sugiyama N.,  2003, \mn@doi [The Astrophysical Journal] {10.1086/375810}, 592, 645

\bibitem[\protect\citeauthoryear{{Yoshida}, {Hosokawa}  \& {Omukai}}{{Yoshida} et~al.}{2012}]{Yoshida12}
{Yoshida} N.,  {Hosokawa} T.,   {Omukai} K.,  2012, \mn@doi [Progress of Theoretical and Experimental Physics] {10.1093/ptep/pts022}, \href {https://ui.adsabs.harvard.edu/abs/2012PTEP.2012aA305Y} {2012, 01A305}

\bibitem[\protect\citeauthoryear{{Zubovas} \& {King}}{{Zubovas} \& {King}}{2021}]{Zubovas21}
{Zubovas} K.,  {King} A.,  2021, \mn@doi [\mnras] {10.1093/mnras/stab004}, \href {https://ui.adsabs.harvard.edu/abs/2021MNRAS.501.4289Z} {501, 4289}

\makeatother
\end{thebibliography}




\FloatBarrier
\appendix

\section{Appendix}
\label{sec:Appendix}
Plots of three quantities (density, temperature, and $X_{\text{H}_2}$) are included for the background-only, short-delay, and long-delay runs. In each plot, we display these quantities at two times. For clump A, the first time shown is the simulation (re)start, snapshot \#10. For clump B, this initial time is snapshot \#17, the first stage with a sink particle in the clump. The second time displayed for both is snapshot \#60. We first display the set of plots for clump B, followed by the plots for clump A.
\label{app:topclump}

\begin{figure*}
\centering
    \includegraphics[width=1\textwidth]{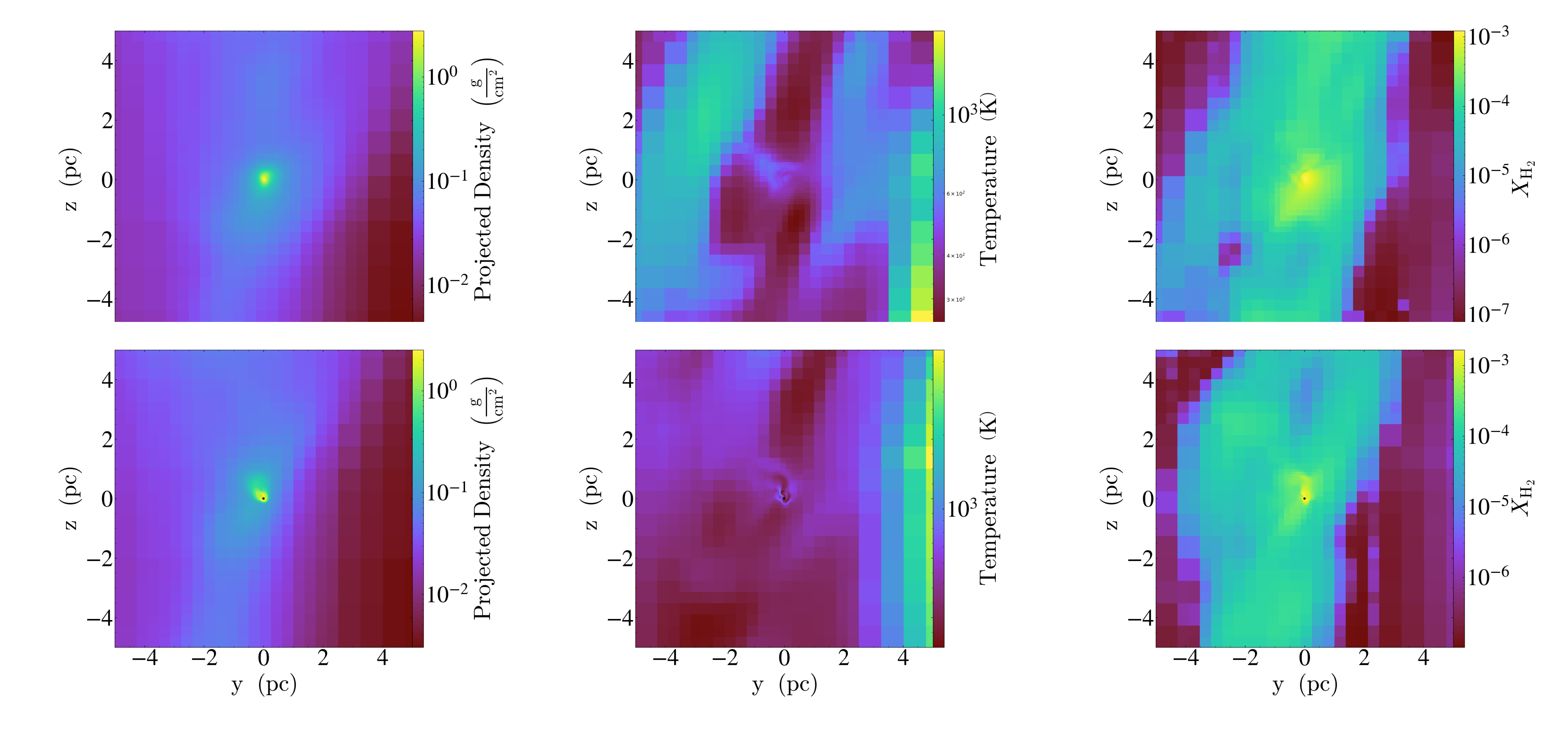}

    \caption{From left to right: projection of density, slice of temperature, and slice of the $X_{\text{H}_2}$ along clump B's x-axis in the background-only LW run. The top row displays these plots at snapshot \#17, $0.07 \, \rm Myr$ after the simulation start, centered about sink B1. The bottom row displays these values at snapshot \#60, $0.43 \, \rm Myr$ later.}
    \label{fig:NoLWBottom}
\end{figure*}

\begin{figure*}
    \centering
    \includegraphics[width=1\textwidth]{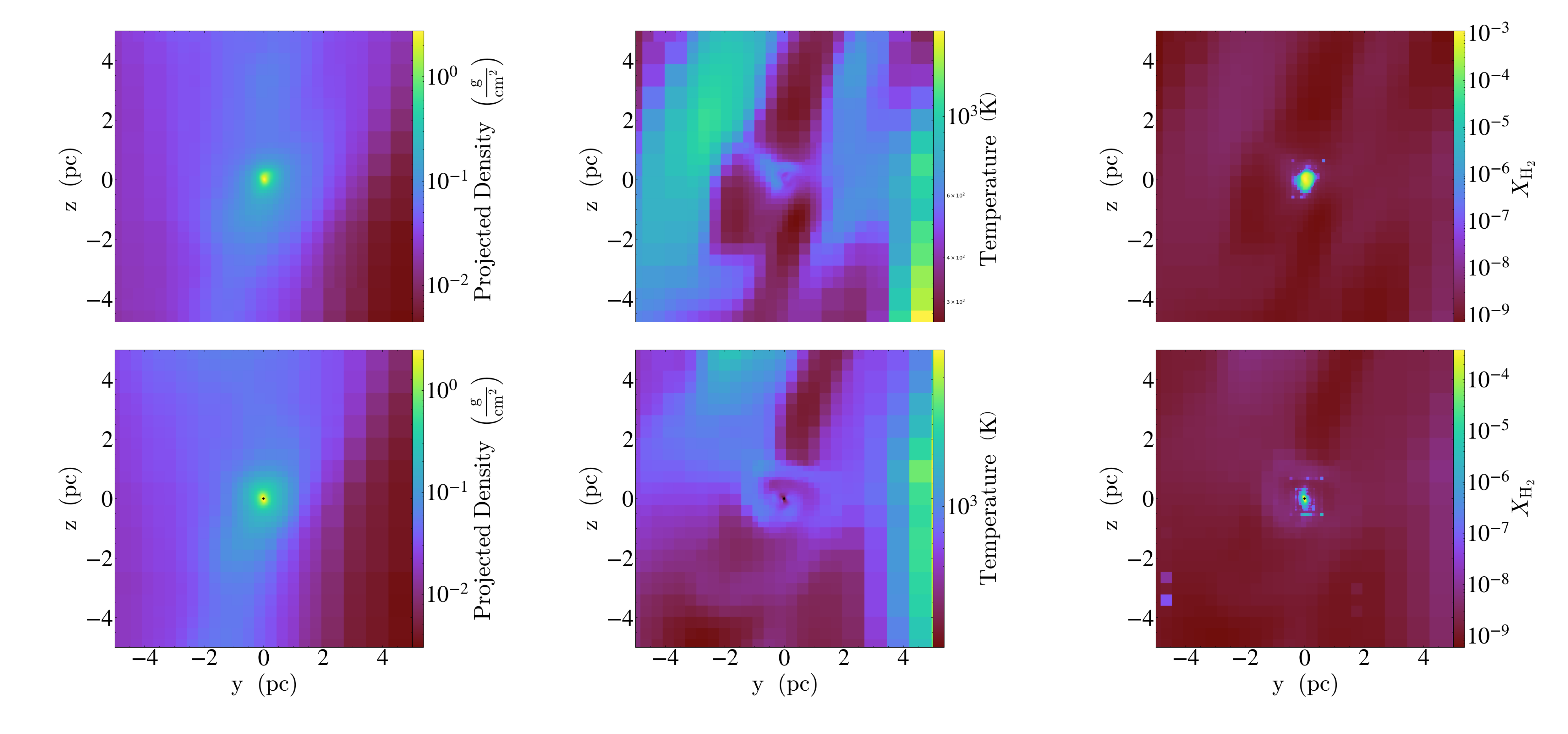}
    
    \caption{Same as Fig.~\ref{fig:NoLWBottom} except displaying the short-delay run.}
    \label{fig:100DelayBottom}
\end{figure*}

\begin{figure*}
    \centering
    \includegraphics[width=1\textwidth]{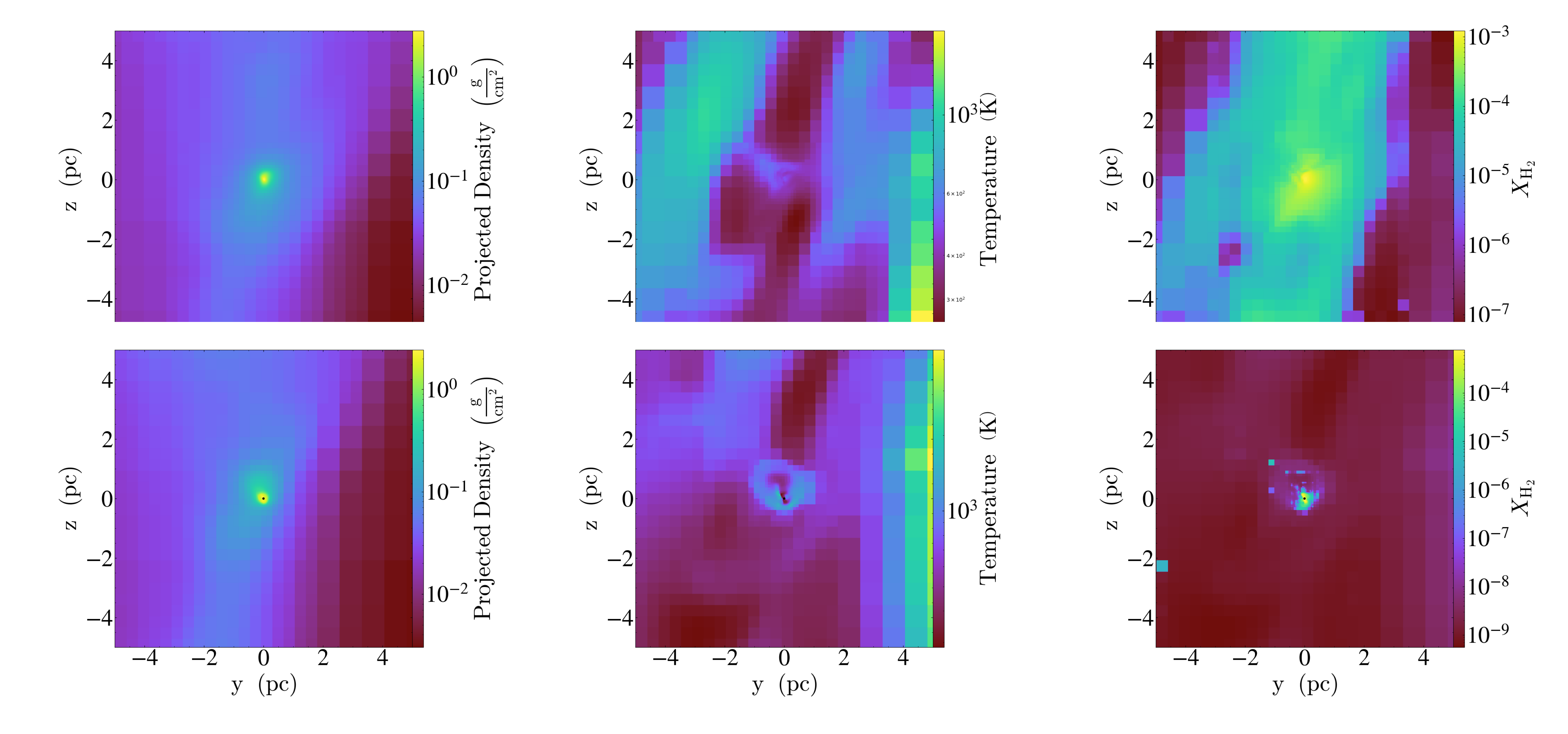}

    \caption{Same as Figs.~\ref{fig:NoLWBottom} and~\ref{fig:100DelayBottom} except displaying the long-delay run.}
    \label{fig:250DelayBottom}
\end{figure*}

\begin{figure*}
    \centering
    \includegraphics[width=1\textwidth]{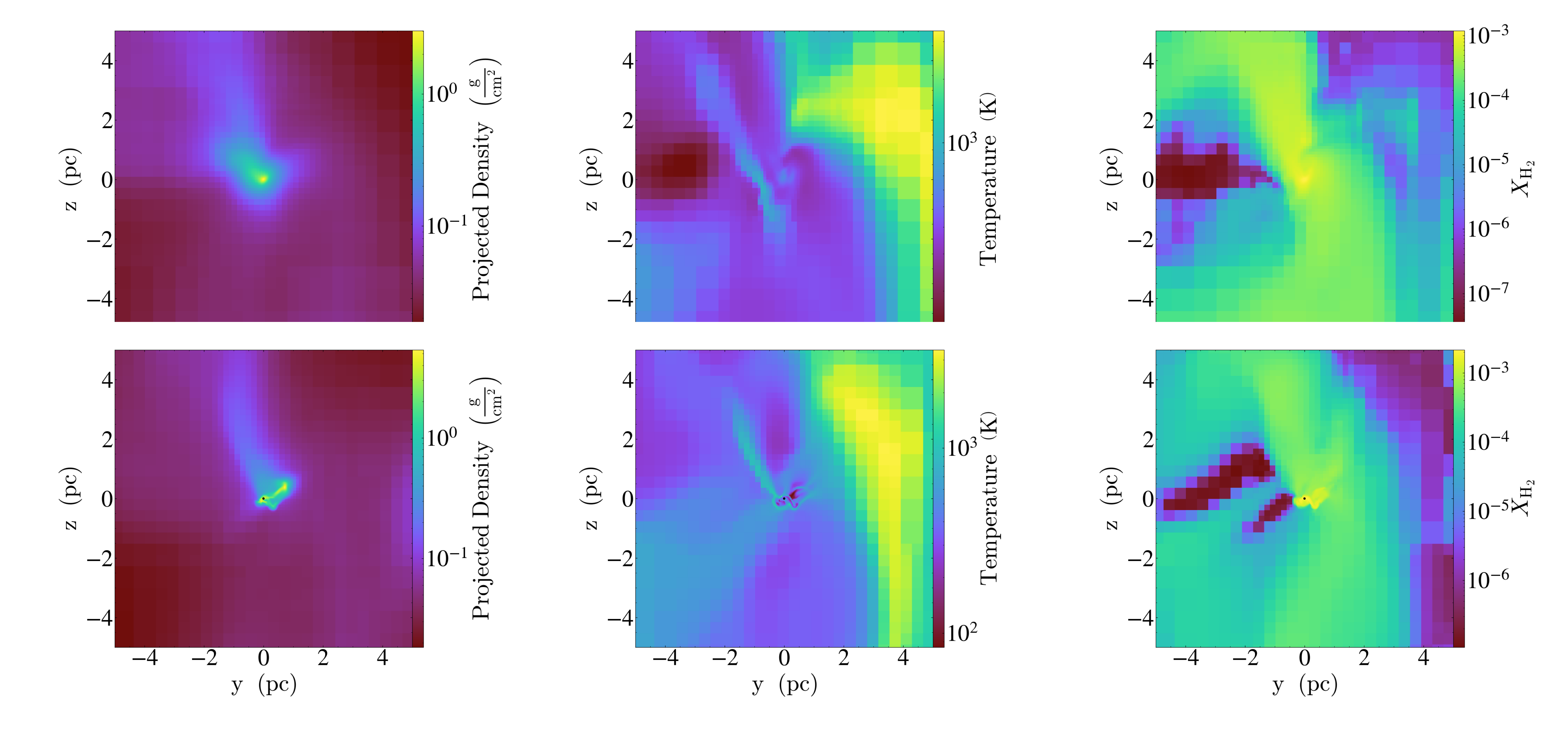}
    
    \caption{Same as Fig.~\ref{fig:NoLWBottom} except centered about sink A1 and clump A. The top row now shows frames from snapshot \#10, the simulation (re)start, centered about clump A's maximum density. The bottom row again displays these values at snapshot \#60, centered about sink A1.}
    \label{fig:NoLWTop}
\end{figure*}

\begin{figure*}
    \centering
    \includegraphics[width=1\textwidth]{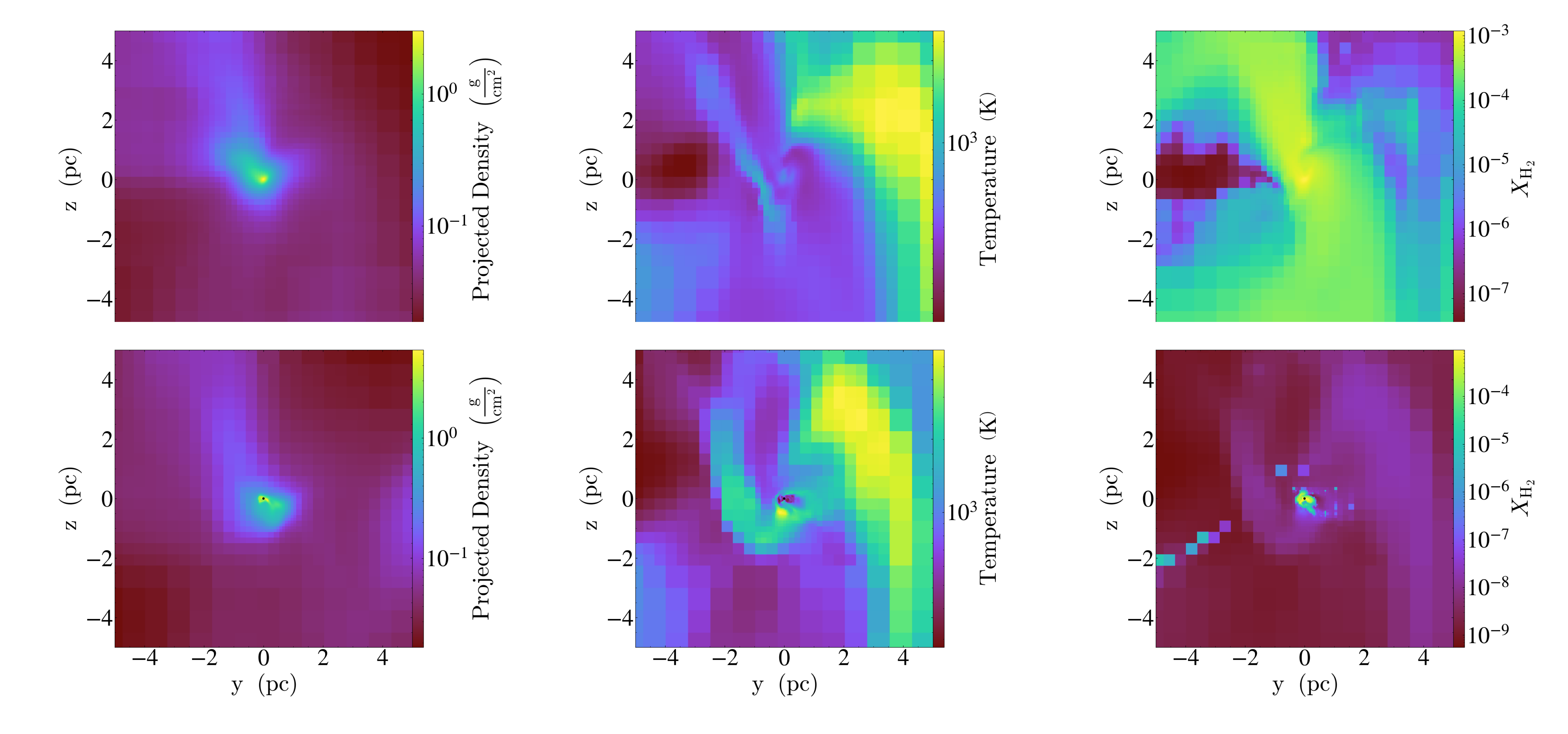}
    
    \caption{Same as Fig.~\ref{fig:NoLWTop} but for the short-delay run.}
    \label{fig:100DelayTop}
\end{figure*}

\begin{figure*}
    \centering
    \includegraphics[width=1\textwidth]{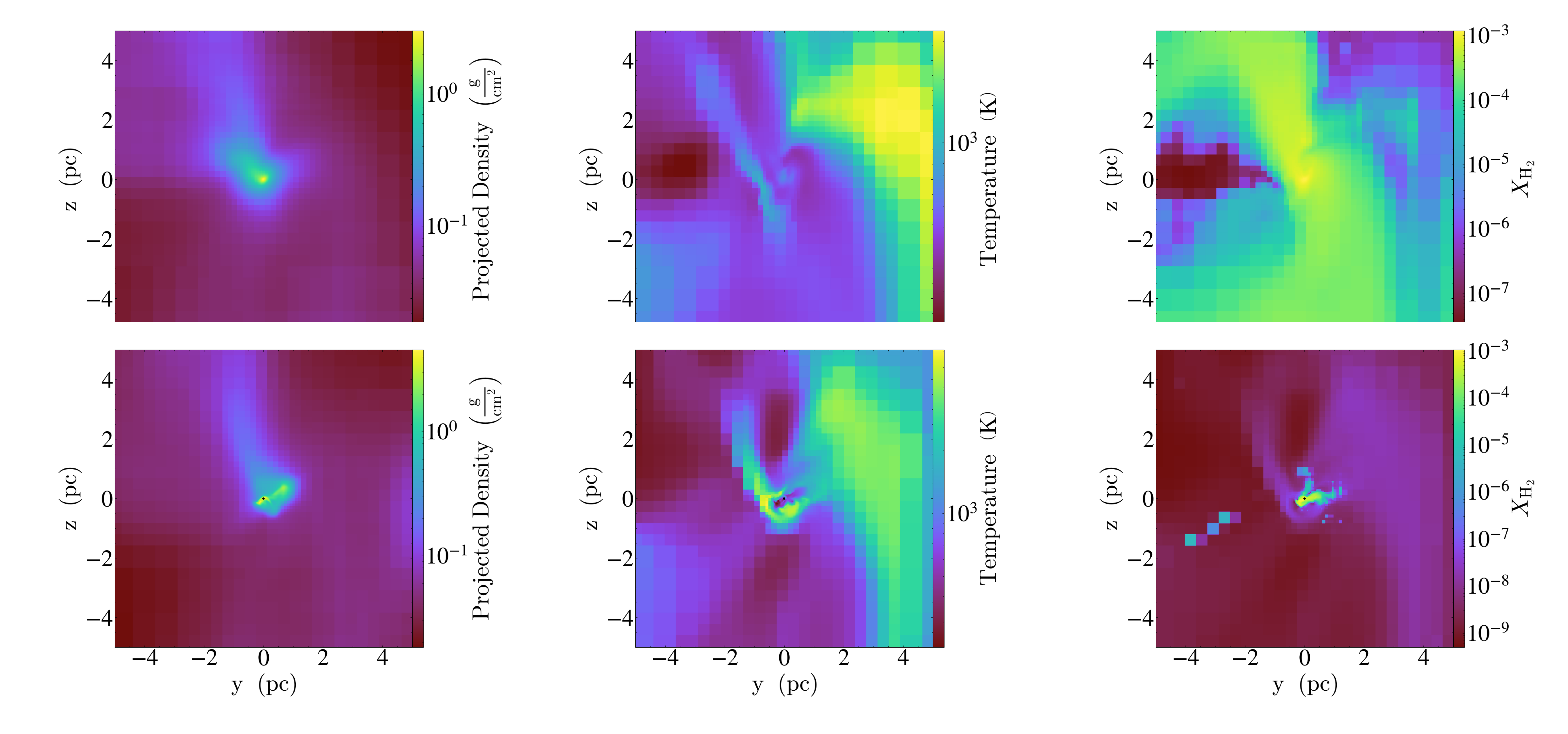}
    
    \caption{Same as Figs.~\ref{fig:NoLWTop} and~\ref{fig:100DelayTop} but for the long-delay run.}
    \label{fig:250DelayTop}
\end{figure*}


\FloatBarrier


\bsp	
\label{lastpage}
\end{document}